\documentclass[pra,twocolumn,showpacs,superscriptaddress,floatfix,amsmath,nofootinbib,amssymb]{revtex4}
\usepackage[english]{babel}
\usepackage{graphicx}% Include figure files
\usepackage{dcolumn}% Align table columns on decimal point
\usepackage{bm}% bold math
\usepackage{verbatim}
\usepackage{mathrsfs}
\usepackage{color}

\newcommand{\ket}[1]{\left| {#1} \right\rangle}
\newcommand{\bra}[1]{\left\langle {#1} \right|}

\newcommand{\proj}[2]{\left| {#1} \right\rangle\!\left\langle {#2} \right|}

\newcommand{\biket}[2]{\left| {#1} \right\rangle_{\text{I}}\left| {#2} \right\rangle_{\text{IV}}}
\newcommand{\bikete}[2]{\left| {#1} \right\rangle_{\text{B}}\left| {#2} \right\rangle_{\bar {\text{B}}}}

\newcommand{\pa}{p}
\newcommand{\tr}{\operatorname{Tr}}
\newcommand{\omegar}{{\omega_\text{R}}}
\newcommand{\diff}{\text{d}}

\def\slashchar#1{\setbox0=\hbox{$#1$} % set a box for #1
\dimen0=\wd0 % and get its size
\setbox1=\hbox{/} \dimen1=\wd1 % get size of /
\ifdim\dimen0>\dimen1 % #1 is bigger
\rlap{\hbox to \dimen0{\hfil/\hfil}} % so center / in box
#1 % and print #1
\else % / is bigger
\rlap{\hbox to \dimen1{\hfil$#1$\hfil}} % so center #1
/ % and print /
\fi}

\begin{document}

%\nofiles
%\preprint{vs.$8.15$}

\title{Unveiling quantum entanglement degradation near a Schwarzschild black hole}% Force line breaks with \\
\author{Eduardo Mart\'{i}n-Mart\'{i}nez}%
 \affiliation{Instituto de F\'{i}sica Fundamental, CSIC, Serrano 113-B, 28006 Madrid, Spain}
%This line break forced with \textbackslash\textbackslash}
 \author{Luis J. Garay}
 %\altaffiliation[Also at ]{Physics Department, XYZ University.}%Lines break automatically or can be forced with \\
\affiliation{Departamento de F\'isica Te\'orica II, Universidad Complutense de Madrid, 28040 Madrid, Spain}
\affiliation{Instituto de Estructura de la Materia, CSIC, Serrano 121, 28006 Madrid, Spain}
%This line break forced with \textbackslash\textbackslash
\author{Juan Le\'on}
 \affiliation{Instituto de F\'{i}sica Fundamental, CSIC, Serrano 113-B, 28006 Madrid, Spain}

\date{June 7, 2010}

\begin{abstract}
We analyze the entanglement degradation provoked by the Hawking
effect in a bipartite system Alice-Rob when Rob is in the proximities of a
Schwarzschild black hole while Alice is free-falling into it. We will obtain the limit in which the tools imported from the Unruh entanglement 
degradation phenomenon can be used properly, keeping control on the
approximation. As a result, we will be able to determine the degree of entanglement as a function
of the distance of Rob to the event horizon, the mass of the black hole,
and the frequency of Rob's entangled modes. By means of this analysis we
will show that all the interesting phenomena occur in the vicinity of the
event horizon and that the presence of event
horizons do not effectively degrade the entanglement when Rob is far off the black hole.
The universality of the phenomenon is presented: There are not fundamental differences
for different masses when working in the natural unit system adapted to each black hole. We also discuss some aspects of the localization of Alice and Rob states. All this study is done without using
the  single mode approximation.
\end{abstract}

\pacs{03.67.Mn, 03.65.-w, 03.65.Yz, 04.62.+v}% PACS, the Physics and Astronomy
                             % Classification Scheme.
%\keywords{Suggested keywords}%Use showkeys class option if keyword
                              %display desired
\maketitle

%\section{\label{sec:level1}First-level heading:\protect\\ The line
%break was forced \lowercase{via} \textbackslash\textbackslash}

%Previous works Rindler

%\cite{DaviesUnr,Unruh,Takagi,Crispino}

%$\mathscr{I}$

\section{Introduction}

Relativistic quantum information is born from the combination of very
different and fruitful branches of physics. Namely, general relativity,
quantum field theory and quantum information theory. Its aim is to study
 the behavior of quantum correlations in relativistic settings. In its scope,
among other topics, is the study of the behavior of quantum correlations in non-inertial settings, which
has produced  abundant literature
\cite{Alsingtelep,TeraUeda2,ShiYu,Alicefalls,AlsingSchul,SchExpandingspace,Adeschul,KBr,LingHeZ,ManSchullBlack,PanBlackHoles,AlsingMcmhMil,DH,Steeg,Edu2,
schacross,Ditta,Hu,DiracDiscord}. This discipline provides novel tools for the
analysis of the Unruh effect and the Hawking effect,  allowing us to
study the behavior of the correlations shared between non-inertial
observers.

In previous works it was analyzed the entanglement degradation
phenomenon produced when one of the partners of an entangled bipartite
system undergoes a constant acceleration; this phenomenon, sometimes
called Unruh decoherence, is strongly related to the Unruh effect. Its
study revealed that there are very strong differences between fermionic
and bosonic field entanglement
\cite{Alicefalls,AlsingSchul,Edu2,Ditta,DiracDiscord}. The reason for these
differences  was traced back to fermionic/bosonic statistics and not to
the difference between bosonic and fermionic mode population as
previously thought \cite{Edu3,Edu4,Edu5}. In these earlier studies some
conclusions were drawn about the infinite acceleration limit, in which the
situation is similar to being arbitrarily close to an event horizon of a
Schwarzschild black hole.

However there are many subtleties and differences between Rindler and
Schwarzschild space-times. For example Schwarzschild space-time
presents a real curvature singularity while Rindler metric is nothing but
the usual Minkowski metric represented in different coordinates and,
therefore, has no singularities. The Rindler horizon is also of very
different nature from the Schwarzschild's event horizon. Namely, the
Rindler horizon is an acceleration horizon experienced only by
accelerated observers (at rest in Rindler coordinates). On the other hand,
a Schwarzschild horizon is an event horizon, which affects the global
causal structure of the whole space-time, independently of the observer.
Also, for the Rindler space-time there are two well defined 
timelike Killing vectors with respect to which modes can be classified
according to the criterion of being of positive or negative frequency.
Contrarily, Schwarzschild space-time has only one timelike Killing vector
(outside the horizon).

Therefore, to analyze the entanglement degradation produced due to the
Hawking effect near a Schwarzschild black hole we must be careful,
above all if we want to do a deeper study than simply taking the limit in
which the Rindler acceleration parameter becomes infinite. In this paper
we will show how we can use the tools coming from the study of the Unruh
degradation in uniformly accelerated scenarios without restricting only
to the exact infinite acceleration limit and controlling to what extent
such tools are valid.

Consequently, we will be able to compute the entanglement degradation
introduced by the Hawking effect as a precise function of three physical
parameters, the distance of Rob to the event horizon, the mass of the
black hole, and the frequency of the mode that Rob has entangled with
Alice's field state. As a result of this study we will obtain not only the
explicit form of the quantum correlations as a function of the physical
parameters mentioned above,  but also a  quantitative control on what
distances from the horizon can be still analyzed using the mathematical
toolbox coming from the Rindler results.

Contrarily to all the previous works in which it was carried out the single
mode approximation described in \cite{AlsingMcmhMil, AlsingSchul} and
of common use in the published literature, we will argue that we do not
need to use such approximation to study the fundamental effects on the
entanglement derived from the Hawking effect.

Our setting consists in two observers (Alice and Rob), one of them
free-falling into a Schwarzschild black hole  close to the horizon (Alice)
and the other one standing at a small distance  from the event horizon
(Rob). Alice and Rob are the observers of a bipartite quantum state which
is maximally entangled for the observer in free fall. The Hawking effect
will introduce degradation in the state as seen by Rob, impeding all the
quantum information tasks between both observers.

In this context we will analyze not only the classical and quantum
correlations between Alice and Rob, but also what is the behavior of the
correlations that both observers would acquire with the mode fields on
the part of the space-time that is classically unaccessible due to the
presence of the event horizon.

By means of this study we will show that all the interesting entanglement
behavior occurs in the vicinity of the event horizon. What is more, we will
argue that as the entangled partners go 
 away from the horizon the effects on entanglement become unnoticeably small and, as a consequence, quantum information
tasks  in universes that contain event horizons are not jeopardized.

We will also show that the phenomenon of the Hawking degradation is
universal for every Schwarzschild black hole, which is to say, it is ruled by
the presence of the event horizon and is not fundamentally influenced by
the specific value of the black hole parameters when the analysis is
performed using natural units to the black hole.  Furthermore, we will discuss the validity of the results obtained when instead of the usual plane wave basis we work in a base of wave packets, for which the states of Alice and Rob can be spatially localized.

%Furthermore,  we will recover all the consequences derived in other studies for the Rindler space-time, so that what was argued before using rough approximations and similitudes, is explicitly obtained here in the Schwarzschild scenario.

%Moreover, this work provides a recipe about how one can study information degradation phenomena in the proximities of black holes using well known tools from the study of uniformly accelerated frames and controlling to what extent such tools are valid.

This paper is structured as follows. In section \ref{sec2} we show how we
work without using the single mode approximation, presenting previous
results about the Unruh entanglement degradation for scalar and Dirac
fields. In section \ref{sec3} we study the entanglement in a
Schwarzschild space-time using the tools built for the Rindler case,
detailing to what extent this approximation holds. In section \ref{sec4}
we will present the result for the correlations between the different
bipartitions in the Schwarzschild space-time scenario. In section \ref{newsec5} we show that the results obtained in the preceding sections are also valid when we consider complete sets of localized modes instead of plane waves bases. Finally, we present
our conclusions in section \ref{conclusions}.

\section{Rindler space-time}\label{sec2}

%\begin{figure}[h]
%\begin{center}
%\includegraphics[width=.75\textwidth]{Rindler}
%\caption{(Colour online) Dirac field: Mutual information tradeoff and conservation law between the systems Alice-Rob and Alice-AntiRob as acceleration varies. It is also shown the behavior of the mutual information for the system Rob-AntiRob. Blue continuous line: Mutual information $AR$, red dotted line: Mutual information $A\bar R$, black dashed line: Mutual information $R\bar R$ }
%\label{Rindler}
%\end{center}
%\end{figure}

Along this work we are going to consider bipartite scalar and Dirac field
states. We will name Alice the observer of the first part of the system
and Rob the observer of the second part. In this fashion, the  quantum
state for the whole system is defined by the tensor product
\begin{eqnarray}
\ket{\phi_\text{A},\phi_\text{R}}\equiv\underbrace{\ket{\phi_\text{A}}}_{\text{Alice's}}\otimes\underbrace{\ket{\phi_\text{R}}}_{\text{Rob's}}.
\end{eqnarray}
Now, while Alice is in an inertial frame, we will consider that Rob is
observing the system from an accelerated frame.

A uniformly accelerated observer viewpoint is described by means of the
well-known Rindler coordinates \cite{gravitation,Takagi}. In order to map
field states in Minkowski space-time to Rindler coordinates, two
different sets of coordinates are necessary. These sets of coordinates
define two causally disconnected regions in Rindler space-time. If we
consider that the uniform acceleration $a$ lies on the $z$ axis, the new
Rindler coordinates $(t,x,y,z)$ as a function of Minkowski coordinates
$(\hat t,\hat x,\hat y,\hat z)$ are
\begin{equation}\label{Rindlcoordreg1}
a\hat t=e^{az}\sinh(at),\; a\hat z=e^{az}\cosh(at),\; \hat x= x,\; \hat y= y
\end{equation}
for region I, and
\begin{equation}\label{Rindlcoordreg2}
a\hat t=-e^{az}\sinh(at),\; a\hat z=-e^{az}\cosh(at),\; \hat x= x,\; \hat y= y
\end{equation}
for region IV.
\begin{figure}
\includegraphics[width=.50\textwidth]{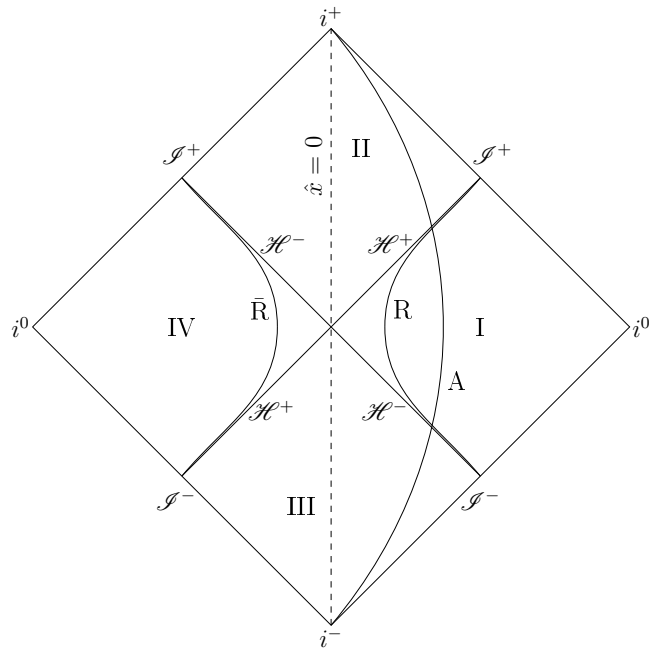}
\caption{Flat space-time conformal diagram showing Alice, Rob and AntiRob
trajectories. $i^0$ denotes the spatial infinities, $i^-$, $i^+$ are
respectively the timelike  past and future infinities,
$\mathscr{I}^-$ and $\mathscr{I}^+$ are the null past and future infinities
respectively, and $\mathscr{H}^\pm$ are the Rindler horizons.}
\label{Rindler}
\end{figure}
As we can see from Fig. \ref{Rindler}, there are two more regions labeled
II and III. To map them we would need to switch
$\cosh\leftrightarrow\sinh$ in Eqns.
\eqref{Rindlcoordreg1} and \eqref{Rindlcoordreg2}. In these regions $t$ is a
spacelike coordinate and $z$ is a timelike coordinate. However, the
solutions of the Klein-Gordon/Dirac equation in such regions are not
required to discuss entanglement between the inertial observer Alice and
the accelerated observer Rob. This is so because Rob would be constrained to either
region I or IV, having no possible access to the opposite regions as they
are causally disconnected
\cite{Birrell,gravitation,Takagi,Alicefalls,AlsingSchul}.

The Rindler coordinates $z,t$ go from $-\infty$ to $\infty$
independently in regions I and IV. Therefore, each region admits a
separate quantization procedure with their corresponding positive and
negative energy solutions of the Klein-Gordon (or Dirac) equations.

The states $\ket{1_{\hat \omega}}_\text{M}=a^\dagger_{\hat
\omega,\text{M}}\ket{0}_\text{M}$ are free massless scalar field
modes, in other words, solutions of positive frequency $\hat\omega$
(with respect to the Minkowski timelike Killing vector $\partial_{\hat
t}$) of the free Klein-Gordon equation:
\begin{align}\label{modmin}
\ket{1_{\hat\omega}}_\text{M}&\equiv u_{\hat\omega}^\text{M}\propto\frac{1}{\sqrt{2{\hat \omega}}}e^{-i {\hat \omega} \hat t},
\nonumber\\
\ket{1_{\hat \omega_1}1_{\hat\omega_2}}&=\ket{1_{\hat\omega_1}}\otimes\ket{1_{\hat\omega_2}},
\end{align}
where only the time dependence has been made explicit. The label
$\text{M}$ just means that these states are expressed in the
Minkowskian Fock space basis.

An accelerated observer   can also define his vacuum and excited states
of the field. Actually, there are two natural vacuum states associated
with the positive frequency modes in regions $\text{I}$ and $\text{IV}$
of Rindler space-time. These are $\ket{0}_\text{I}$ and
$\ket{0}_{\text{IV}}$, and subsequently we can define the field
excitations using Rindler coordinates $(x,t)$ as
\begin{align}\label{modrin}
\ket{1_\omega}_\text{I}&=a^\dagger_{\omega,\text{I}}\ket{0}_\text{I}
\equiv u_\omega^\text{I}\propto\frac{1}{\sqrt{2\omega}}e^{-i\omega  t},\nonumber\\
\ket{1_\omega}_{\text{IV}}&=a^\dagger_{\omega,\text{IV}}\ket{0}_\text{IV}
\equiv u_\omega^\text{IV}\propto\frac{1}{\sqrt{2\omega}}e^{i\omega  t}.
\end{align}
These modes are related by a space-time reflection and only have support in regions I and IV of the
Rindler space-time respectively.

However, these Rindler modes  are not independent of the Minkowskian
modes. Indeed we can expand the field in terms of Minkowski modes and,
independently, in terms of Rindler modes. Therefore, the Minkowskian 
and the Rindler set of modes are related by a change of basis
\cite{Edu2,Birrell}. The relationship between the Minkowski Fock basis
and the two Rindler Fock bases comes through the Bogoliubov
coefficients, obtained by equating the field expansion in Minkowskian
modes with the field expansion in Rindler modes. In general we have that
\begin{equation}
u_{\hat\omega_j}^\text{M}=\sum_i\left(\alpha^\text{I}_{ij}u_{\omega_i}^\text{I}+
\beta^{\text{IV}*}_{ij}u_{\omega_i}^{\text{IV}*}+\alpha^\text{IV}_{ij}
u_{\omega_i}^\text{IV}+\beta^{\text{I}*}_{ij}u_{\omega_i}^{\text{I}*}\right).
\end{equation}
The Bogoliubov coefficient matrices $\alpha^R_{ij}, \beta^{R}_{ij}$
(where $R=\text{I},\text{IV}$) are given by the Klein-Gordon scalar
product between both sets of modes
\begin{equation}
\alpha^{R}_{ij}=\left(u_{\omega_i}^R,u_{\hat\omega_j}^\text{M}\right),
\qquad\beta^{R}_{ij}=-\left(u_{\omega_i}^{R},u_{\hat\omega_j}^{\text{M}*}\right).
\end{equation}

The relationship between modes also establishes a relationship between
the Minkowski annihilation operator   and the particle operators in Rindler
regions I and IV:
\begin{equation}
a_{\hat\omega_j,\text{M}}\!=\!\sum_i\left(\alpha^{\text{I}*}_{ij}
a^{\phantom{\dagger}}_{\omega_i,\text{I}}\!+\!\beta^{\text{IV}}_{ij}
a^\dagger_{\omega_i,\text{IV}}+\alpha^{\text{IV}*}_{ij}
a^{\phantom{\dagger}}_{\omega_i,\text{IV}}\!+\!\beta^{\text{I}}_{ij}
a^\dagger_{\omega_i,\text{I}}\right).
\end{equation}

On the other hand there exist an infinite number of orthonormal bases
that define the same vacuum state, namely the Minkowski vacuum
$\ket{0}_\text{M}$, which can be used to expand the solutions of the
Klein-Gordon equation. More explicitly, since the modes
$u^\text{M}_{\hat\omega_i}$ have positive frequency, any complete set
made out of independent linear combinations of these modes only
(without including the negative frequency ones
$u^{\text{M}*}_{\hat\omega_i}$) will define the same vacuum
$\ket{0}_\text{M}$.

Specifically, as described in e.g. Refs. \cite{Unruh,Takagi,Birrell} and
explicitly constructed below, there exists an orthonormal basis
$\{\psi_{\omega_j}^\text{M}, \psi'^{\text{M}}_{\omega_j}\}$ determined by certain linear
combinations of monochromatic positive frequency modes,
$u^\text{M}_{\hat\omega_i}$
\begin{equation}\label{modopsi}
\psi^\text{M}_{\omega_j}=\sum_i C_{ij}\, u_{\hat\omega_i}^\text{M},\qquad \psi'^{\text{M}}_{\omega_j}=\sum_i C'_{ij}\, u_{\hat\omega_i}^\text{M},
\end{equation}
 such that the Bogoliubov coefficients that relate this basis
 $\{\psi_{\omega_j}^\text{M},\psi'^{\text{M}}_{\omega_j}\}$  and the Rindler basis
 $\{u^\text{I}_{\omega_i},u^\text{IV}_{\omega_i}\}$ have the following form:
 \begin{align}\label{bogo1}
\hat\alpha^{\text{I}}_{ij}
&=\left(u_{\omega_i}^\text{I},\psi_{\omega_j}^\text{M}\right)=\cosh r_{\text{s},i}\,\delta_{ij},
\nonumber\\
\hat\alpha^{\text{IV}}_{ij}
&=\left(u_{\omega_i}^{\text{IV}},\psi_{\omega_j}^\text{M}\right)=0,\nonumber\\
\hat\beta^{\text{I}}_{ij}
&=-\left(u_{\omega_i}^\text{I},\psi_{\omega_j}^{\text{M}*}\right) = 0,
\nonumber\\
\hat\beta^{\text{IV}}_{ij}&=-\left(u_{\omega_i}^{\text{IV}},\psi_{\omega_j}^{\text{M}*}\right)=- \sinh r_{\text{s},i}\,\delta_{ij},
\end{align}
and analogously for $\hat\alpha'^{\text{I,IV}}_{ij}$ and $\hat\beta'^{\text{I,IV}}_{ij}$ interchanging the labels $\text{I}$ and $\text{IV}$ in the formulas above. In this expressions
\begin{equation}\label{defr1}
\tanh r_{\text{s},i}=\exp(-\pi {\omega_i }/{a}),
\end{equation}
and the label $\text{s}$ in $r_{\text{s},i}$ has been introduced to
indicate that we are dealing with a scalar field. In this expression and in
what follows, we will use Planck units ($\hbar=c=G=1$).

In this fashion a mode $\psi_{\omega_j}^\text{M}$ (or a mode $\psi'^{\text{M}}_{\omega_j}$) expands only in
terms of mode of frequency $\omega_j$ in Rindler regions
$\text{I}$ and $\text{IV}$ and for this reason we have labeled $\psi_{\omega_j}^\text{M}$ and $\psi'^{\text{M}}_{\omega_j}$  with the frequency $\omega_j$ of the corresponding Rindler modes. In other words, we can express a given
monochromatic Rindler mode of frequency $\omega_j$   as a linear
superposition of the single Minkowski modes $\psi_{\omega_j}^{\text{M}}$ and $\psi'^{\text{M}*}_{\omega_j}$ or as a polychromatic combination of the
positive frequency Minkowski modes $u^\text{M}_{\hat\omega_i}$ and
their conjugates.

Let us denote ${a_{\omega_j}}$ and $a_{\omega_j}^\dagger$ the
annihilation and creation operators associated with modes
$\psi^\text{M}_{\omega_j}$ (analogously we denote ${a'_{\omega_j}}$ and $a_{\omega_j}'^\dagger$ the ones associated with modes $\psi'^{\text{M}}_{\omega_i}$). The Minkowski vacuum
$\ket0_\text{M}$, which is annihilated by all the Minkowskian operators
$a_{\hat \omega_i,\text{M}}$, is also annihilated by all the operators
$a_{\omega_j}$ and $a_{\omega_j}'$, as we already mentioned. This comes out because any
combination of Minkowski annihilation operators annihilates the
Minkowskian vacuum.

Due to the Bogoliubov relationships \eqref{bogo1} being diagonal, each annihilation
operator $a_{\omega_i}$ can be expressed as a combination of Rindler
particle operators of only one Rindler frequency  $\omega_i$:
\begin{equation}\label{buenmod}
a_{\omega_i}=\cosh r_{\text{s},i}\, a^{\phantom{\dagger}}_{\omega_i,\text{I}}-
\sinh r_{\text{s},i}\,a^\dagger_{\omega_i,\text{IV}},
\end{equation}
and analogously for $a'_{\omega_i}$ interchanging the labels I and IV.

An analogous procedure can be carried out for fermionic fields (e.g. Dirac
fields). We can use linear combinations of monochromatic solutions of the
Dirac equation $\psi^\text{M}_{\omega_i,\sigma}$ and $\bar
\psi^\text{M}_{\omega_i,\sigma}$ (and their primed versions) built in the same fashion as for scalar
fields:
%\begin{align}\label{modopsif}
%\psi^\text{M}_{\omega_j,\sigma}=\sum_i D_{ij}\, u_{\hat\omega_i,\sigma}^\text{M},
%& \bar \psi^\text{M}_{\omega_j,\sigma}=\sum_i E_{ij}\, v_{\hat\omega_i,\sigma}^\text{M},\\
%\psi'^{\text{M}}_{\omega_j,\sigma}=\sum_i D'_{ij}\, u_{\hat\omega_i,\sigma}^\text{M},
%& \bar\psi'^{\text{M}_{\omega_j,\sigma}=\sum_i E'_{ij}\, v_{\hat\omega_i,\sigma}^\text{M},
%\end{align}
\begin{align}\label{modopsif}
\psi^\text{M}_{\omega_j,\sigma}=\sum_i D_{ij}\, u_{\hat\omega_i,\sigma}^\text{M},
&\qquad \bar \psi^\text{M}_{\omega_j,\sigma}=\sum_i E_{ij}\, v_{\hat\omega_i,\sigma}^\text{M},\nonumber\\
\psi'^{\text{M}}_{\omega_j,\sigma}=\sum_i D'_{ij}\, u_{\hat\omega_i,\sigma}^\text{M},
& \qquad\bar\psi'^{\text{M}}_{\omega_j,\sigma}=\sum_i E'_{ij}\, v_{\hat\omega_i,\sigma}^\text{M},
\end{align}
where $u_{\hat\omega_i,\sigma}^\text{M}$ and
$v_{\hat\omega_i,\sigma}^\text{M}$ are respectively monochromatic
solutions of positive (particle) and negative (antiparticle) frequency $\pm\hat\omega_i$ of the
massless Dirac equation with respect to the Minkowski Killing time. The
label $\sigma$ accounts for the possible spin degree of freedom of the
fermionic field\footnote{Throughout this work we will consider that the
spin of each mode is in the acceleration direction and, hence, spin will not
undergo Thomas precession due to instant Wigner rotations
\cite{AlsingSchul,Jauregui}.}.

The coefficients of these combinations are such that for the modes
$\psi_{\omega_i,\sigma}$ and $\bar \psi_{\omega_i,\sigma}$ the
annihilation operators are related with the Rindler ones by means of the
following Bogoliubov transformations \cite{Jauregui,chinada,AlsingSchul}:
\begin{eqnarray}\label{Bogoferm}
\nonumber c_{{\omega_i},\sigma}&=&\cos{r_{\text{d},i}}\,c_{\text{I},{\omega_i},
\sigma}-\sin r_{\text{d},i}\,d^\dagger_{\text{IV},{\omega_i},-\sigma},\\*
d_{{\omega_i},\sigma}^\dagger&=&\cos{r_{\text{d},i}}\,d^\dagger_{\text{IV},
{\omega_i},\sigma}+\sin r_{\text{d},i}\,c_{\text{I},{\omega_i},-\sigma},
\end{eqnarray}
and analogously for $c'_{\omega_i}$ and $d'^\dagger_{\omega_i}$ interchanging the labels I and IV, where
\begin{equation}\label{defr}
\tan r_{\text{d},i}=\exp(-\pi {{\omega_i}}/{a}).
\end{equation}
Here $c_{{\omega_i},\sigma}$, $d_{{\omega_i},\sigma}$ represent the
annihilation operators of modes $\psi^\text{M}_{\omega_i,\sigma}$ and
$\bar
\psi^\text{M}_{\omega_i,\sigma}$ for particles and antiparticles respectively. 
The label $\text{d}$ in $r_{\text{d},i}$ has been introduced to indicate
Dirac field. The specific form for $\psi^\text{M}_{\omega_i,\sigma}$
and $\bar
\psi^\text{M}_{\omega_i,\sigma}$ as a linear combination of monochromatic
solutions of the Dirac equation can be seen, for instance, in
\cite{Jauregui,chinada} among many other references. Notice again that,
although we are denotating $a_{\omega_i},
c_{\omega_i,\sigma},d_{\omega_i,\sigma}$ the operators associated
with Minkowskian modes, those modes are not monochromatic, but a
linear combination of monochromatic modes given by \eqref{modopsi} and
\eqref{modopsif}.

As we are going to discuss fundamental issues and not an specific
experiment, there is no reason to adhere to a specific basis. Specifically,
if we work in the bases \eqref{modopsi},Ê\eqref{modopsif} for Minkowskian modes we do
not need to carry out the single mode approximation
\cite{AlsingMcmhMil,AlsingSchul} in which one single mode   of Minkowski
frequency $\hat\omega_i$ was expressed as a monochromatic
combination of Rindler modes of the same frequency. This approximation
has allowed pioneering studies of correlations with non inertial observers,
but it is based on misleading assumptions on the characteristics of Rob's
detector, being partially flawed. In any case, discussing the validity of
the single mode approximation is not the aim of this work. A complete
discussion of the problems associated with the single mode approximation
and how to overcome them is in course of completion and will be reported
elsewhere \cite{estos}. As far as this work is concerned it is enough to
say that we are not using any similar approximation.

\subsection{Vacuum and first excitation for a scalar field}

The Minkowski vacuum state of the field $\ket{0}_\text{M}$ is
annihilated by the annihilation operators $a_{\hat\omega_i,\text{M}}$ as well as
by the operators $a_{\omega_i}$ and $a'_{\omega_i}$. For the excited states of the field,
we will work with the orthonormal basis $\{\psi^\text{M}_{\omega_i},\psi'^{\text{M}}_{\omega_i}\}$
defined in \eqref{modopsi} such that
\begin{equation}\label{onepartgoodb}
\ket{1_{\omega_i}}_\text{M}=a_{\omega_i}^\dagger \ket{0}_\text{M},\qquad |1'_{\omega_i}\rangle_\text{M}=a'^\dagger_{\omega_i} \ket{0}_\text{M}
\end{equation}
are solutions of the free Klein-Gordon equation which are not
monochromatic, but linear superpositions of plane waves of positive
frequency $\hat\omega_j$.

As shown in \cite{Takagi,Birrell}, we can express the Minkowski vacuum
state in terms of the Rindler Fock space basis,
$\ket{0}_\text{M}=\bigotimes_{i}\ket{0_{\omega_i}}_\text{M}$,
where
\begin{equation}\label{scavacinf1}
\ket{0_{\omega_i}}_\text{M}=\frac{1}{\cosh r_{\text{s},i}}\sum_{n=0}^\infty
(\tanh r_{\text{s},i})^n \ket{n_{\omega_i}}_\text{I}\ket{n_{\omega_i}}_{\text{IV}}.
\end{equation}
It is straightforward to check that this vacuum is, indeed, annihilated by the
operators $a_{\omega_i}$ and $a'_{\omega_i}$.

The Minkowskian one particle state $\ket{1_{\omega_i}}_\text{M}$  (in the basis
\eqref{modopsi}) results from applying the creation operator
$a^\dagger_{\omega_i}$ to the vacuum state. We can also translate it
to the Rindler basis
\begin{align}\label{unoinf1}
\ket{1_{\omega_i}}_\text{M}&=\frac{1}{(\cosh r_{\text{s},i})^2}\nonumber\\
&\times
\sum_{n=0}^{\infty}  (\tanh  r_{\text{s},i})^n
\sqrt{n+1}\ket{n+1_{\omega_i}}_\text{I} \ket{n_{\omega_i}}_{\text{IV}}.
\end{align}
The mode $ |1'_{\omega_i}\rangle_\text{M}$ is analogous but swapping the labels I and IV.

\subsection{Vacuum and first excitation for a Dirac field}

For simplicity in the notation we are going to present only the construction of the modes $\psi_{\omega_i}^\text{M}$ and $\bar \psi_{\omega_i}^\text{M}$ in \eqref{modopsi} since, as we will show later when we build the entangled state, they are the only ones of importance in this analysis. In any case, the construction of the primed modes for the Dirac field is analogous, but we would need to expand the notation below to include the possibility of antiparticles in Rindler region I and particles in Rindler region IV being careful with all the anticommutation subtleties typical for fermionic fields.

As for the scalar case, the vacuum state of the field
$\ket{0}_\text{M}$ is annihilated by the annihilation operators
$c_{\hat\omega_{i},\sigma,\text{M}}$ and $d_{\hat\omega_{i},\sigma,\text{M}}$ for all
$\hat\omega_i,\sigma$ as well as by the operators
$c_{\omega_i,\sigma}$ and $d_{\omega_i,\sigma}$ for all
$\omega_i,\sigma$.

For the excited states of the field, we will work with the orthonormal
basis \eqref{modopsif} such that
\begin{equation}\label{onepartgood}
\ket{\sigma_{\omega_i}}_\text{M}=c_{\omega_i,\sigma}^\dagger \ket{0}_\text{M}
\end{equation}
are positive frequency solutions of the free Dirac equation which are not
monochromatic, but linear superpositions of plane waves of positive
frequency $\hat\omega_i$.
%\begin{equation}
%\ket{1_{{\omega_i}}}_\text{M}\equiv {\psi^\text{M}_i}\propto \sum_j C_{ij}\frac{1}{\sqrt{2\omega}}e^{-i \omega_j \hat t}
%\end{equation}
%\begin{equation}
%\ket{1_{\omega_1}1_{\omega_2}}=\ket{1_{\omega_1}}\otimes\ket{1_{\omega_2}}
%\end{equation}
%The label $M$ just means that those states are expressed in the Minkowskian Fock space basis PAGER.

Let us introduce some notation for the Rindler field excitations that will follow the same
convention as in \cite{Edu2,Edu3,Edu4}. 
\begin{align}\label{notation2}
 \ket{\sigma_{{\omega_i}}}_\text{I}
 &=c^\dagger_{\text{I},{\omega_i},\sigma}\ket{0}_\text{I},\nonumber\\
  \ket{\sigma_{{\omega_i}}}_\text{IV}
  &=d^\dagger_{\text{IV},{\omega_i},\sigma}\ket{0}_{\text{IV}},\nonumber\\
 \ket{\pa_{{\omega_i}}}_\text{I}
 &=c^\dagger_{\text{I},{\omega_i},\uparrow}
 c^\dagger_{\text{I},{\omega_i},\downarrow}\ket{0}_\text{I}
 =-c^\dagger_{\text{I},{\omega_i},\downarrow}c^\dagger_{\text{I},{\omega_i},
 \uparrow}\ket{0}_\text{I},\nonumber\\
\ket{\pa_{{\omega_i}}}_{\text{IV}}
&=d^\dagger_{\text{IV},{\omega_i},\uparrow}d^\dagger_{\text{IV},{\omega_i},
\downarrow}\!\ket{0}_{\text{IV}}=-d^\dagger_{\text{IV},{\omega_i},\downarrow}
d^\dagger_{\text{IV},{\omega_i},\uparrow}\!\ket{0}_{\text{IV}},
\end{align}
where $p_{{\omega_i}}$ represents the spin pair state in the mode with
frequency ${\omega_i}$. Notice that due to the operator
anticommutation relations,
\begin{align}\label{notation3}
\nonumber \ket{\sigma_{{\omega_i}}}_\text{I}
\ket{\sigma'_{{\omega_i}}}_{\text{IV}}
&=c^\dagger_{\text{I},{{\omega_i}},\sigma}
d^\dagger_{\text{IV},{\omega_i},\sigma'}
\ket{0}_{\text{I}}\ket{0}_{\text{IV}}\nonumber\\
&=-d^\dagger_{\text{IV},{{\omega_i}},\sigma'}
c^\dagger_{\text{I},{\omega_i},\sigma}\ket{0}_{\text{I}}\ket{0}_{\text{IV}},\nonumber\\
d^\dagger_{\text{IV},{{\omega_i}},\sigma'}\biket{\sigma_{{\omega_i}}}{0}&=-\biket{\sigma_{{\omega_i}}}{\sigma'_{{\omega_i}}}.
\end{align}

As it can be seen in \cite{Edu2}, the projection onto the unprimed sector of the basis \eqref{modopsif} of the Minkowski vacuum state written in the
Rindler basis, is as follows
\begin{align}\label{vacuumf1}
 \ket{0_{{\omega_i}}}_\text{M}&=(\cos r_{\text{d},i})^2\biket{0}{0}
 \nonumber\\
 &+
 \sin r_{\text{d},i} \cos r_{\text{d},i}
 \left(\biket{\uparrow_{{\omega_i}}}{\downarrow_{{\omega_i}}}
+ \biket{\downarrow_{{\omega_i}}}{\uparrow_{{\omega_i}}}\right)\nonumber\\
&+
(\sin r_{\text{d},i})^2\biket{\pa_{{\omega_i}}}{\pa_{{\omega_i}}}.
\end{align}
It is straightforward to check that the vacuum is annihilated by
$c_{{\omega_i},\sigma}$ and $d_{{\omega_i},\sigma}$ simply using
\eqref{Bogoferm} and applying both operators to \eqref{vacuumf1}.

The one particle state (projected onto the sector $\psi^\text{M}_{\omega_i}$ of \eqref{modopsif}) in the Rindler basis can be readily obtained by
applying the particle creation operator $c^\dagger_{{\omega_i},\sigma}$ to \eqref{vacuumf1}:
\begin{align}\label{onepartf1}
\ket{\uparrow_{{\omega_i}}}_\text{M}&=\cos r_{\text{d},i}
\biket{\uparrow_{{\omega_i}}}{0}+
\sin r_{\text{d},i}\biket{\pa_{{\omega_i}}}{\uparrow_{{\omega_i}}},\nonumber\\
\ket{\downarrow_{{\omega_i}}}_\text{M}&=\cos r_{\text{d},i}
\biket{\downarrow_{{\omega_i}}}{0}-
\sin r_{\text{d},i}\biket{\pa_{{\omega_i}}}{\downarrow_{{\omega_i}}}.
\end{align}

\subsection{Entanglement degradation due to Unruh effect}

We will now summarize the results that have been obtained concerning
the effects of an uniform acceleration on quantum correlations.

Let us first consider the following maximally entangled state for a scalar
field:
\begin{equation}\label{Min1}
\ket{\Psi}_\text{s}=\frac{1}{\sqrt2}\left(\ket{0}_\text{A}\ket{0}_\text{R}
+\ket{1}_\text{A}\ket{1_{\omegar}}_\text{R}\right),
\end{equation}
where the label A denotes Alice's subsystem and R denotes Rob's
subsystem. In this expression, $\ket{0}_{\text{A,R}}$ represents the
Minkowski vacuum for Alice and Rob, $\ket{1}_\text{A}$ is an arbitrary
one particle state excited from the Minkowski vacuum for Alice, and the
one particle state for Rob is expressed in the basis~\eqref{modopsi} and
characterized by the frequency $\omegar$ observed by Rob.

The election of the modes $\ket{1_{\omega_\text{R}}}$ instead of $|1'_{\omega_\text{R}}\rangle$ to build the maximally entangled state is not relevant since choosing the primed modes would just be equivalent to say that Rob is in region IV instead of in region I, and there is complete symmetry in all the analysis for both cases.

Since the second partner (Rob) ---who  observes the bipartite state
\eqref{Min1}--- is accelerated, it is convenient to map the second partition
of this state  into the Rindler Fock space basis, which can be computed
using equations
\eqref{scavacinf1} and \eqref{unoinf1}, and rewrite  it
in the standard language of relativistic quantum information (i.e., naming
Alice to the Minkowskian observer, Rob to a hypothetic observer in
Rindler's region I and AntiRob to a hypothetic observer in Rindler Region
IV):
\begin{align}\label{Minscalar}
\ket\Psi_\text{s}&=\sum_{n=0}^\infty \frac{(\tanh r_{\text{s}})^n}{\sqrt{2}
\cosh r_{\text{s}}} \bigg(\ket{0}_\text{A}
\ket{n_{\omegar}}_\text{R}\ket{n_{\omegar}}_{\bar{\text{R}}}\nonumber\\
&+\left.\frac{\sqrt{n+1}}{\cosh r_{\text{s}}}\ket{1}_\text{A}
 \ket{n+1_{\omegar}}_\text{R}\ket{n_{\omegar}}_{\bar{\text{R}}}\right),
 \end{align}
where
\begin{equation}
\tanh r_\text{s}=\exp(-\pi\omegar/a).
\end{equation}

The same can be done in the case of a Dirac field. Let us now consider the
following maximally entangled state for a Dirac field in the Minkowskian
basis
\begin{equation}\label{Minf}
\ket{\Psi}_\text{d}=\frac{1}{\sqrt2}\left(\ket{0}_\text{A}\ket{0}_\text{R}+
\ket{\uparrow}_\text{A}\ket{\downarrow_\omegar}_\text{R}\right).
\end{equation}
As for the bosonic case, if Rob, who observes this bipartite state, is
accelerated, it is convenient to map the second partition of this state into
the Rindler Fock space basis, which can be computed using Eqns.
\eqref{vacuumf1} and \eqref{onepartf1}. The explicit form of such state
can be seen in \cite{Edu2}.

Notice that we have chosen a specific maximally entangled state
\eqref{Minf} of all the possible choices. This election has no relevance
since in \cite{Edu3} it was shown the universality of the degradation of
fermionic entanglement. All fermionic maximally entangled states are
equally degraded by the Unruh effect, no matter what kind of maximally
entangled state is (either occupation number or spin Bell state), or even if
we work with a Grassmann scalar field instead of a Dirac field.

Let us denote
\begin{equation}\label{dens}
\rho^\text{s}_{\text{AR}\bar{\text{R}}}=\proj{\Psi_\text{s}}{\Psi_\text{s}},\qquad \rho^\text{d}_{\text{AR}\bar{\text{R}}}=\proj{\Psi_\text{d}}{\Psi_\text{d}},
\end{equation}
the tripartite density matrices for the bosonic and fermionic cases,  in
which we use the Minkowski basis for Alice and the Rindler basis for
Rob-AntiRob. One could ask what is the physical meaning  of each of
these three `observers'. Alice represents an observer in an inertial frame.
For Alice the states
\eqref{Min1} and \eqref{Minf} are maximally entangled.
Rob represents an accelerated observer moving in a $x=a^{-1}$
trajectory in Region I of Rindler space-time (as seen in Fig. \ref{Rindler}) who shares a bipartite entangled state \eqref{Min1} or \eqref{Minf} with Alice.
AntiRob represents an observer  moving in a $x=a^{-1}$ trajectory in
Region IV with access to the information to which Rob is not able to
access (at least classically) due to the presence of the Rindler horizon.

In the standard Unruh entanglement degradation scenario
\cite{Alicefalls,Edu2}, as Rob is not able to access AntiRob part of the
system we must trace over AntiRob degrees of freedom when accounting
for the quantum state shared by Alice and Rob. This provokes, for
instance, the observation of a thermal bath by Rob while Alice observes
the Minkowski vacuum as it can be seen elsewhere
\cite{Birrell,AlsingSchul,Edu2}. As a consequence the state becomes
mixed, which causes some degree of correlation loss in the system AR as
we increase the value of the acceleration $a$. In references
\cite{Alicefalls,AlsingSchul,Edu2,Edu3,Ditta,Edu4,Edu5,DiracDiscord} it is
studied how this phenomenon affects the entanglement for different
fields.

It has been also studied \cite{AlsingSchul,Edu4,Edu5} the correlation
trade-off among the all possible bipartitions of the system, namely,
 Alice-Rob (AR),  Alice-AntiRob $(\text{A}\bar{\text{R}})$, and
  Rob-AntiRob $(\text{R} \bar{\text{R}})$.

Bipartition AR is the most commonly considered in the literature. It
represents the system formed by an inertial observer and the modes of
the field which an accelerated observer is able to access. The second
bipartition $(\text{A}\bar{\text{R}})$ represents the subsystem
formed by the inertial observer Alice and the modes of the field which
Rob is not able to access due to the presence of an horizon as he
accelerates. Classical communication between the two partners is only
allowed for the bipartitions AR and $\text{A}\bar{\text{R}}$. We will
call these bipartitions  `Classical communication allowed' (CCA) from now
on. These bipartitions are the only ones in which quantum information
tasks are possible to be performed.

On the other hand, no quantum information tasks can be performed using
$\text{R}\bar{\text{R}}$ correlations since classical communication
between Rob and AntiRob is not allowed. Anyway, studying this
bipartition is still necessary to give a complete description of the
behavior of the correlation created between the space-time regions
separated by the horizon.

As it is commonplace in quantum information, the partial quantum states
for each bipartition are obtained by tracing over the third subsystem
\begin{align}\label{traza1}
\nonumber\rho^{\text{AR}}& = \tr_{\bar{\text{R}}}\rho^{\text{AR}\bar{\text{R}}},\\
\nonumber\rho^{\text{A}\bar{\text{R}}}& = \tr_{\text{R}}\rho^{\text{AR}\bar{\text{R}}},\\
\rho^{\text{R}\bar{\text{R}}}&=\tr_{\text{A}}\rho^{\text{AR}\bar{\text{R}}}.
\end{align}

In the cases AR and $\text{A}\bar{\text{R}}$, there are physical
arguments to justify the need for this `tracing over' beyond mere
quantum information considerations, namely, Rob will never be  able to
access region IV of the space-time due to the presence of the Rindler
horizon so that $\bar{\text{R}}$ (Region $\text{IV}$) must be traced
out. Likewise, AntiRob is not able to access region I  because of the
horizon and hence R (Region I) must be traced out. For the subsystem
${\text{R}\bar{\text{R}}}$ this tracing over subsystem A corresponds
to the standard procedure for analyzing correlations between two parts
of a multipartite system. The properties of the correlations among these
subsystems has been analyzed in the literature, showing a completely
different behavior of quantum correlations for the CCA bipartitions
depending on whether the system is fermionic or bosonic.

For fermionic fields, quantum correlations are conserved as Rob
accelerates \cite{AlsingSchul,Edu3}. Specifically, as entanglement in the
bipartition AR is reduced, entanglement in the system
$\text{A}\bar{\text{R}}$ is increased. In the limit of
$a\rightarrow\infty$ some entanglement survives in all the bipartitions of
the system.

For the scalar field the situation is radically different, namely, no
entanglement is created in the CCA bipartitions. Moreover, the
entanglement in the AR bipartition is very quickly lost as Rob accelerates,
even if we artificially limit the dimension of the Hilbert space
\cite{Edu5}.

This different behavior, thought at the beginning to be a consequence of
the finite dimensionality of the fermionic Hilbert space, was
demonstrated to be ruled only by statistics \cite{Edu3,Edu4,Edu5}, which
plays a crucial role in the phenomenon of Unruh entanglement
degradation. The role of statistics is so important that, for fermions, the
behavior of quantum correlations has been proven to be universal
\cite{Edu3}. Also, the survival of entanglement for the fermionic case, is
arguably related to statistical correlations \cite{Edu3,Edu4}. All these
aspects will be discussed in depth later on, when we present the results
for the Schwarzschild black hole.

\section{The ``Black Hole Limit'': translation Rindler-Kruskal}\label{sec3}

In this section we will study a completely new setting using the tools
learned from \cite{Alicefalls,AlsingSchul,Edu3,Edu4}. We will prove in a
constructive way that the entanglement degradation in the vicinity of an
eternal black hole can be studied in detail with these well-known tools. By
means of the construction shown below we will be able to deal with new
problems such as computing entanglement loss between a free-falling
observer and another one placed at fixed distance to the event horizon as
a function of the distance, studying the behavior of quantum
correlations in the presence of black holes. We will also show that the
entanglement loss produced by an eternal black hole shows universality.

To begin this section let us work a little bit with the Schwarzschild
metric
 \begin{equation}
 \diff s^2=-\left(1-\frac{2m}{r}\right)\diff t^2+\left(1-\frac{2m}{r}\right)^{-1}
 \diff r^2 +r^2\diff \Omega^2,
 \end{equation}
 where $m$ is the black hole mass and $\diff\Omega^2$ is the line element in the unit sphere.
Due to the symmetry of the problem we are going to restrict the analysis
to the radial coordinate. To shorten notation let us write the radial part
of metric as
 \begin{equation}\label{propsch}
 \diff s^2=-f\diff t^2+f^{-1}\diff r^2,
 \end{equation}
 where $f=1-2m/r$.

 We can choose to write the metric in terms of the proper time $t_0$ of an observer placed in $r=r_0$ as follows,
  \begin{equation}\label{propsch2}
 \diff s^2=-\frac{f}{f_0}\diff t_0^2+f^{-1}\diff r^2,
 \end{equation}
where $f_0=1-2m/r_0$.
The relationship between $t_0$ and $t$ is given by the norm of the
timelike Killing vector $\xi=\partial_t$ in $r=r_0$, namely
 $t_0=\sqrt{f_0}\,t$.

We can now change the spatial coordinate such that the new coordinate
vanishes at the Schwarzschild radius $r=R_{\text{S}}=2m$. Let us define
$z$ in the following way
 \begin{equation}\label{defz}
 r-2m=\frac{z^2}{8m}\quad\Rightarrow\quad
  f=\frac{(\kappa z)^2}{1+(\kappa z)^2},
 \end{equation}
 with $\kappa =1/(4m)$ being the surface gravity of the black hole. Then the metric \eqref{propsch2} results
 \begin{equation}\label{change1}
 \diff s^2=-\frac{1}{f_0}\frac{(\kappa z)^2}{1+(\kappa z)^2}\diff t_0^2+
 \left[1+(\kappa z)^2\right]\diff z^2.
 \end{equation}
Near the event horizon ($z\approx0$), we can expand this metric to
lowest order in $z$ and approximate it by
%\begin{equation}{1+(\kappa z)^2}\approx 1\end{equation}
%Therefore, the metric can be approximated in this region by
\begin{equation}\label{misli}
\diff s^2=-\left(\frac{\kappa z}{\sqrt{f_0}}\right)^2\diff t_0^2+\diff z^2,
\end{equation}
which is a Rindler metric with acceleration parameter
$\kappa/\sqrt{f_0}$.

  On the other hand, Eq. \eqref{misli} represents
the metric near the event horizon in terms of the proper time of an
observer placed at $r=r_0$. The next step is giving a physical meaning to
this Rindler-like acceleration parameter. For this, we need to compute the
proper acceleration of a Schwarzschild observer placed at $r=r_0$,
which is, indeed, different from $\kappa$ (as $\kappa$ would be the
acceleration of an observer arbitrarily close to the horizon as seen from
a free-falling frame).

To compute $a$ for this observer as seen by himself (proper
acceleration) we must start from the Schwarzschild metric. The value of
the proper acceleration for an accelerated observer at arbitrary fixed
position $r$ is $a=\sqrt{a_\mu a^\mu}$ where $a^\mu=v^\nu\nabla_\nu
v^\mu$ is the observer 4-acceleration at such position, whereas $v^\mu$
is his 4-velocity.

The 4-velocity for a Schwarzschild observer in an arbitrary position $r$
is
\begin{equation}
v^\mu= {\xi^\mu}/{|\xi|},
\end{equation}
where $\xi\equiv\partial_t$ is the Schwarzschild timelike Killing vector.
As $\xi^\mu=(1,0,0,0)$ in Schwarzschild coordinates, then
$|\xi|=\sqrt{|g_{00}|}=\sqrt{f}$, and therefore
$v^\mu=\xi^\mu/\sqrt{f}$. Thus, we can compute the acceleration
4-vector
\begin{equation}
a^\mu=v^\nu\nabla_\nu v^\mu=\frac{1}{|\xi|}\xi^\nu\nabla_\nu
 \frac{\xi^\mu}{|\xi|} .
\end{equation}
Taking into account that $\xi^\mu$ is a Killing vector and, therefore, it
satisfies $\nabla_\mu\xi_\nu+\nabla_\nu\xi_\mu=0$, we easily obtain
\begin{equation}
a_\mu=\frac12\frac{\partial_\mu|\xi|^2}{|\xi|^2} =
 \frac{\partial_\mu f}{2f}=\frac{1}{2f}\left(0,\partial_r f ,0,0\right).
\end{equation}
Hence, since $g^{rr}=f$, the   proper acceleration for this observer is
\begin{equation}\label{properpropera}
a=\sqrt{g^{\mu\nu}a_\mu a_\nu}=\sqrt{\frac{(\partial_r f)^2}{4f} } .
\end{equation}
For an observer placed at $r=r_0$,
\begin{equation}\label{properpropera3}
a_0=\frac{\kappa}{\sqrt{f_0}}(1-f_0)^2.
\end{equation}

We know from \eqref{defz} that $1-f_0=[1+(\kappa z_0)^2]^{-1}$. So,
if the observer in $r=r_0$ is close to the event horizon ($r_0\approx
R_{\text{S}}$), then, to lowest order, $1+(\kappa z_0)^2\approx 1$ and
\begin{equation}\label{properproperaf}
a\approx {\kappa}/{\sqrt{f_0}}.
\end{equation}
Therefore, under this approximation, we can re-write \eqref{misli} as
\begin{equation}\label{Rindleradapt}
\diff s^2=-\left(a_0 z\right)^2\diff t_0^2+\diff z^2.
\end{equation}
This shows that the Schwarzschild metric can be approximated, in the
proximities of the event horizon, by a Rindler metric whose acceleration
parameter is the proper acceleration of an observer resisting in a position
$r_0$ close enough to the event horizon.

This approximation holds if
\begin{equation}\label{conditioaproximatio}
\left(\frac{z_0}{2R_{\text{S}}}\right)^2\ll1
\end{equation}
or, in other words, if
\begin{equation}\label{conditioaproximatio2}
\frac{\Delta_0}{R_\text{S}}\ll1,
\end{equation}
where $\Delta_0\equiv r_0-R_\text{S}$ is the distance from $r_0$ to
the event horizon. In the limit $r_0\rightarrow R_{\text{S}}$ we obtain
that $f_0\rightarrow0$ and, from \eqref{properproperaf},
$a_0\rightarrow\infty$. This shows rigorously that being very close to
the event horizon of a Schwarzschild black hole can be very well
approximated by the infinite acceleration Rindler case, as it was
suggested in \cite{Alicefalls,AlsingSchul,Edu3,Edu4}. This also enables us
to study what would happen  with the entanglement between observers
placed at different distances of the event horizon as far as the Rindler
approximation holds.

\begin{figure}
\begin{center}
\includegraphics[width=.50\textwidth]{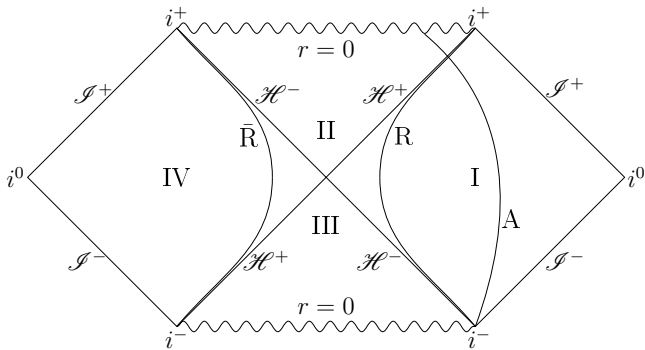}
\caption{Kruskal space-time  conformal diagram showing trajectories for Alice,
Rob and AntiRob. $i^0$ denotes the spatial infinities, $i^-$, $i^+$
are respectively the timelike past and future infinities,
$\mathscr{I}^-$ and $\mathscr{I}^+$ are the null past and future infinities
respectively, and $\mathscr{H}^\pm$ are the event horizons.}
\label{Kruskal}
\end{center}
\end{figure}

Now let us identify again who is who in this new scenario. For this, we
introduce the null Kruskal-Szeckeres coordinates
\begin{equation}\label{eq:szeckeres}
u=-\kappa^{-1}\exp[-\kappa (t-r^*)],\quad v=\kappa^{-1}\exp[\kappa (t+r^*)],
\end{equation}
where $r^*=r+2m\log|1-r/2m|$. In terms of these coordinates the radial
part of the Schwarzschild metric is
\begin{equation}
\diff s^2=\frac{-1}{2\kappa r}e^{-2\kappa r}\diff u\diff v,
\end{equation}
where $r$ is implicitly defined by (\ref{eq:szeckeres}). The Penrose
diagram for this maximal analytic extension is shown in Fig.~\ref{Kruskal}. In this coordinates, 
near the horizon the metric can be written to  lowest order as
\begin{equation}
\diff s^2=-e^{-1}\diff u \diff v
\end{equation}
and $uv=-(\kappa z)^2$.

Hence, there are three regions in which we can clearly define physical
timelike vectors respect to which we can classify positive and negative
frequencies:
\begin{itemize}
\item  $\partial_{\hat t}\propto (\partial_{u}+\partial_{v})$.
The parameter $\hat t$ for this timelike vector corresponds to the
proper time of a free-falling observer close to the horizon, and it is
analogous to the Minkowskian   timelike Killing vector. Positive
frequency modes associated to this timelike vector define a vacuum
state known as the Hartle-Hawking vacuum $\ket{0}_\text{H}$,
which is analogous to $\ket{0}_\text{M}$ in the Rindler case.

\item $\partial_{t}\propto (u\partial_{u}-v\partial_{v})$.
 It is the Schwarzschild timelike Killing vector, which (when properly
normalized) corresponds to an observer whose acceleration at the
horizon equals the surface gravity $\kappa$ of the black hole with
respect to a Minkowskian observer, or, in other words, with proper
acceleration $a_0\approx\kappa/\sqrt{f_0}$ close to the horizon.
The vacuum state corresponding to positive frequencies associated to
this timelike Killing vector is called the Boulware vacuum
$\ket{0}_\text{B}$. This state is analogous to the Rindler vacuum
$\ket{0}_\text{I}$.

\item There is another timelike Killing vector $-\partial_{t}$
(as in Rindler) for region IV that will allow us to define another
Boulware vacuum  in region IV. We will call it AntiBoulware vacuum
$\ket0_{\bar{\text{B}}}$, analogous to $\ket{0}_\text{IV}$ in the
Rindler case.
\end{itemize}

Now, in this scenario,
$\ket{1_{\hat\omega}}_\text{H}=a^\dagger_{\hat\omega,\text{H}}\ket{0}_\text{H}$
are free scalar field modes, in other words, solutions of positive
frequency ${\hat\omega}$ with respect to $\partial_{\hat t}$  of the
free Klein-Gordon equation close to the horizon
\begin{equation}
\ket{1_{\hat\omega}}_\text{H}\equiv u_{\hat\omega}^\text{H}\propto
\frac{1}{\sqrt{2\hat\omega}}e^{-i\hat\omega \hat t}.
\end{equation}
The label H just means that those states are expressed in the
Hartle-Hawking Fock space basis.

An observer located at a fixed distance from the black hole can also define his own vacuum and
excited states of frequency $\omega$ respect to the Killing vector
$\partial_{t}$. Actually, there are two natural vacuum states associated
with the positive frequency modes in both sides of the horizon these are
$\ket{0}_\text{B}$ and $\ket{0}_{\bar{\text{B}}}$, vacua for the
positive frequency modes in regions I and IV respectively (Fig.
\ref{Kruskal}). Subsequently, for a scalar field, we can define the field
excitations as
\begin{align}
\ket{1_\omega}_{\text{B}}&=a^\dagger_{\omega,\text{B}}\ket{0}_\text{B}\equiv
u_\omega^{\text{B}}\propto\frac{1}{\sqrt{2\omega}}e^{-i\omega  t},\nonumber\\
\ket{1_\omega}_{\bar{\text{B}}}&=a^\dagger_{\omega,\bar{\text{B}}}
\ket{0}_{\bar{\text{B}}}\equiv u_\omega^{\bar{\text{B}}}\propto\frac{1}{\sqrt{2\omega}}e^{i\omega  t}.
\end{align}

%reeescribir
Then, the analogy between the Rindler-Minkowski and the Boulware-Hartle-Hawking states,  and their relation with 
the standard Alice-Rob-AntiRob notation is as follows:
\begin{equation}\label{identif}
\begin{array}{lclcl}
\ket{0}_\text{R}&\leftrightarrow&\ket{0}_\text{I}&\leftrightarrow&\ket{0}_\text{B},\\
\ket{0}_{\bar{\text{R}}}&\leftrightarrow&\ket{0}_{\text{IV}}&
\leftrightarrow&\ket{0}_{\bar{\text{B}}},\\
\ket{0}_{\text{A}}&\leftrightarrow&\ket{0}_\text{M}&\leftrightarrow&
\ket{0}_{\text{H}}.
\end{array}
\end{equation}
The change of basis between Hartle-Hawking modes
and Boulware modes is completely analogous to the change of basis
between Minkowskian modes and Rindler modes with an acceleration
parameter $a_0=\kappa/\sqrt{f_0}$.

In the same fashion as for Rindler we define an orthonormal basis 
%\begin{equation}\label{HHs}
%\psi^\text{H}_{\omega_j}=\sum_i C_{ij}\, u^{\text{H}}_{\hat\omega_i}\qquad \psi'^{\text{H}}_{\omega_j}=\sum_i C'_{ij}\, u^{\text{H}}_{\hat\omega_i}\
%\end{equation}
of Hartle-Hawking scalar field modes $\{\psi^\text{H}_{\omega_j},\psi'^{\text{H}}_{\omega_j}\}$ whose elements are superpositions of  positive-frequency solutions
$u_{\hat\omega_i}^\text{H}$ of the Klein-Gordon equation with
respect to the Kruskal time $\hat t$
 such that each element  corresponds to Boulware modes of one single frequency in the Kruskal regions I and IV
($u^\text{B}_{\omega_j}$ and $u^{\bar{\text{B}}*}_{\omega_j}$). The same can be
done for the Dirac field.
%\begin{equation}\label{HHf}
%\psi^\text{H}_{\omega_j,\sigma}=\sum_i D_{ij}\, u^{\text{H}}_{\hat\omega_i,\sigma}
%\qquad \bar \psi^\text{H}_{ \omega_j,\sigma}=\sum_i E_{ij}\,
%v^{\text{H}}_{\hat\omega_i,\sigma}.
%\end{equation}

We can express the Hartle-Hawking vacuum state in terms of the
Boulware Fock space basis.  To do so we use what we learned from the
Rindler case. Taking into account that $\ket{0}_\text{H}=\bigotimes_{
i}\ket{0_{\omega_i}}_\text{H}$, we have that
\begin{equation}\label{scavacinf}
\ket{0_{\omega_i}}_\text{H}=\frac{1}{\cosh q_{\text{s},i}}
\sum_{n=0}^\infty (\tanh q_{\text{s},i})^n  \ket{n_{\omega_i}}_\text{B}
\ket{n_{\omega_i}}_{{\bar{\text{B}}}},
\end{equation}
where
\begin{equation}\label{defr3}
\tanh q_{\text{s},i}=\exp\left({-\pi \sqrt{f_0}\,\omega_i/{\kappa}}\right)
%=\exp\left[{-4\pi \omegar \sqrt{m^2-\dfrac{2m^3}{r_0}}}\right]
.
\end{equation}
The unprimed Hartle-Hawking one particle state in the basis $\{\psi^\text{H}_{\omega_j},\psi'^{\text{H}}_{\omega_j}\}$ results
from applying the corresponding creation operator to the vacuum state.
We can also translate this state to the Boulware basis:
  \begin{align}\label{unoinf}
\ket{1_{\omega_i}}_\text{H} &= \frac{1}{(\cosh q_{\text{s},i})^2}\nonumber\\
&\times
\sum_{n=0}^{\infty} (\tanh  q_{\text{s},i})^n\sqrt{n+1}
\ket{n+1_{\omega_i}}_\text{B}\!\ket{n_{ \omega_i}}_{{\bar{\text{B}}}}.
\end{align} 

The Hartle-Hawking  vacuum (projected onto the unprimed sector) for the Dirac case is expressed in the
Boulware basis as follows
\begin{align}\label{vacuumf}
 \ket{0_{\omega_i}}_{\text{H}} &= (\cos q_{\text{d},i})^2
 \bikete{0_{\omega_i}}{0_{\omega_i}}\nonumber\\
 &
 +\sin q_{\text{d},i}\cos q_{\text{d},i}
 \left(\ket{\uparrow_{\omega_i}}_{{\text{B}}}
  \ket{\downarrow_{\omega_i}}_{\bar{\text{B}}}+
 \bikete{\downarrow_{\omega_i}}{\uparrow_{\omega_i}}\right)
 \nonumber\\
 &+(\sin q_{\text{d},i})^2\bikete{\pa_{\omega_i}}{\pa_{\omega_i}},
\end{align}
whereas the projected Hartle-Hawking one particle state 
is expressed in the Boulware basis as
\begin{align}\label{onepartf}
 \ket{\uparrow_{\omega_i}}_\text{H}&= \cos q_{\text{d},i}
 \bikete{\uparrow_{\omega_i}}{0_{\omega_i}}+
 \sin q_{\text{d},i}\bikete{\pa_{\omega_i}}{\uparrow_{\omega_i}},\nonumber\\
\ket{\downarrow_{\omega_i}}_\text{H}&=
\cos q_{\text{d},i} \bikete{\downarrow_{\omega_i}}{0_{\omega_i}}-
\sin q_{\text{d},i}\bikete{\pa_{\omega_i}}{\downarrow_{\omega_i}},
\end{align}
where this time
\begin{equation}\label{defr4}
\tan q_{\text{d},i}=\exp\left(-\pi \sqrt{f_0}\,\omega_i /{\kappa}\right)
%=\exp\left[{-4\pi \omegar \sqrt{m^2-\dfrac{2m^3}{r_0}}}\right]
.
\end{equation}

Thus, in this new scenario, we can consider a bipartite state analogous to
the states \eqref{Min1} and \eqref{Minf} for the Rindler scenario which looks
like follows, for fermions and bosons, in the basis of a free-falling observer (Alice)
\begin{align}\label{Haw1}
\ket\Psi_\text{s}&=\frac{1}{\sqrt{2}}
\left(\ket{0}_\text{A}\ket{0}_\text{R}+
\ket{1}_\text{A}\ket{1_{\omegar}}_\text{R}\right),\\
\label{Hawf}
\ket{\Psi}_\text{d}&=\frac{1}{\sqrt2}\left(\ket{0}_\text{A}
\ket{0}_\text{R}+\ket{\uparrow}_\text{A}
\ket{\downarrow_{\omegar}}_\text{R}\right).
\end{align}
This bipartite system consists in two subsystems, the first one is going to
be observed by Alice, who is free-falling into the black hole and close to
the event horizon, and the second one will be observed by Rob, who is near
the event horizon at $r=r_0\approx R_\text{S}$. Therefore, the second
partner who observes the bipartite states
\eqref{Haw1} and \eqref{Hawf}  describes them using the Boulware
basis, so that it is convenient to map the second partition of these states
into the Boulware Fock space basis.

Following the notation \eqref{identif}, to analyze the correlations among
the bipartite subsystems we need to trace out the third subsystem
analogously to what we did in \eqref{traza1}:
\begin{eqnarray}\label{traza2}
\nonumber\rho^{\text{AR}}&\!\!=\!&\tr_{\bar{\text{R}}}\rho^{\text{AR}\bar{\text{R}}},\\*
\nonumber\rho^{\text{A}\bar{\text{R}}}&\!\!=\!&\tr_{\text{R}}\rho^{\text{AR}\bar{\text{R}}},\\*
\rho^{\text{R}\bar{\text{R}}}&\!\!=\!&\tr_{\text{A}}\rho^{\text{AR}\bar{\text{R}}}.
\end{eqnarray}

It can be seen in Fig. \ref{Kruskal} that all the information beyond the
event horizon cannot be accessed by Rob. Actually, what happens beyond
the horizon is determined by the information that Rob can access along
with the information that AntiRob can access. In this context it makes
sense to say that studying the system $\rho^{\text{R}\bar{\text{R}}}$
gives an idea of the correlations across the horizon.

\section{Correlations behavior}\label{sec4}

In this section we will use the machinery we already have from the Rindler
set-ups to compute the entanglement degradation as a function of the
position of Rob.

First we will consider that Rob's frequency $\omegar$ is measured in
natural units adapted to each black hole. This will show how modes of
different frequencies  suffer different correlation degradation. It will
also show how less massive black holes produce a higher degradation than
the heavier ones. Furthermore, this analysis will show the universality of
the phenomenon of the Hawking entanglement degradation for
Schwarzschild black holes.

After that, we will  analyze the different degree of entanglement
degradation experimented by an observer of fixed Boulware frequency
$\omegar$ standing at fixed distances from the event horizon for
different black hole masses.

In the following subsections we will see that all the interesting behavior
happens in regions in which the Rindler approximation
\eqref{Rindleradapt} is valid. Specifically, in the plots below, the
values of the distance to the horizon from which the interesting
entanglement behavior appears are $\Delta_0\lesssim 0.05R_\text{S}$
in all the cases considered in this section for which, consequently, the
approximation \eqref{Rindleradapt} holds.

\subsection{Adapted frequency}

In terms of the mode frequency measured by Rob  (written in units
natural to the black hole, i.e. in terms of the surface gravity $\kappa$) and his position measured in Schwarzschild
radii,
\begin{align}
\Omega&=2\pi\omegar
/\kappa=8\pi m\omegar,\\
R_0&=r_0/R_\text{S}=r_0/(2m),
\end{align}
 Eqns. \eqref{defr3} and
\eqref{defr4} can be written as
\begin{align}\label{defr32}
\tanh q_\text{s}&=\exp\left({-\frac{\Omega}{2} \sqrt{1-\frac{1}{R_0}}}\right),
\\
\label{defr42}
\tan q_\text{d}&=\exp\left({-\frac{\Omega}{2} \sqrt{1-\frac{1}{R_0}}}\right),
\end{align}
showing that the phenomenon of Hawking entanglement degradation
presents universality, which is to say, if the frequency is measured in natural
units, every Schwarzschild black hole behaves in the same way, as
expected.

\subsubsection{Quantum correlations}

We will use the negativity ($\mathcal{N}$) to account for the quantum
correlations between the different bipartitions of the system. It is an
entanglement monotone sensitive to distillable entanglement. The
negativity is defined as the sum of the negative eigenvalues of the partial
transpose density matrix for the system, which is defined as the
transpose of only one of the subsystem q-dits in the bipartite density
matrix. For a general density matrix of a bipartite system AB,
\begin{equation}
\rho_{\text{AB}}=\sum_{ijkl}\rho_{ijkl}\ket{i}_\text{A}\ket{j}_\text{B}\bra{k}_\text{A}\bra{l}_\text{B},
\end{equation}
the partial transpose is defined as
\begin{equation}
\rho_{\text{AB}}^\text{pT}=\sum_{ijkl}\rho_{ijkl}\ket{i}_{\text{A}}\ket{l}_\text{B}\bra{k}_\text{A}\bra{j}_\text{B}.
\end{equation}
 If $\lambda_i$ are the eigenvalues of $\rho^\text{pT}_{\text{AB}}$, then
\begin{equation}\label{negativitydef}
\mathcal{N}_{\text{AB}}=\frac12\sum_{ i}(|\lambda_i|-\lambda_i)=-\sum_{\lambda_i<0}\lambda_i.
\end{equation}

\begin{figure}
\begin{center}
\includegraphics[width=.50\textwidth]{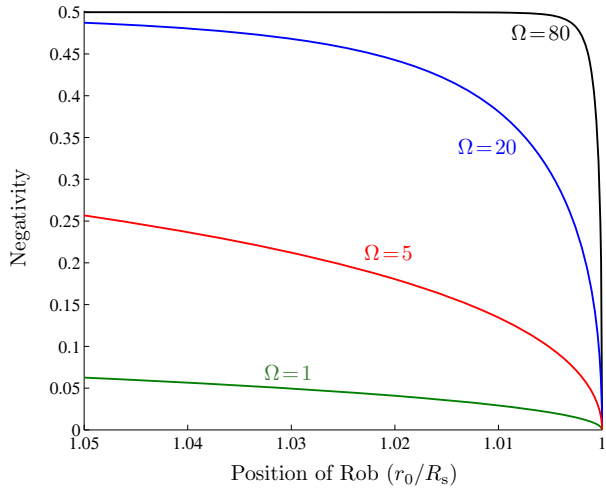}
\caption{Scalar field: Entanglement of the system Alice-Rob as a function of the position of Rob for different values of $\Omega$. Entanglement vanishes as Rob approaches
the Schwarzschild radius while no entanglement is created between Alice and AntiRob. The smaller the value of $\Omega$ the more degradation is produced by the black hole.}
\label{sncca}
\end{center}
\end{figure}

\begin{figure}
\begin{center}
\includegraphics[width=.50\textwidth]{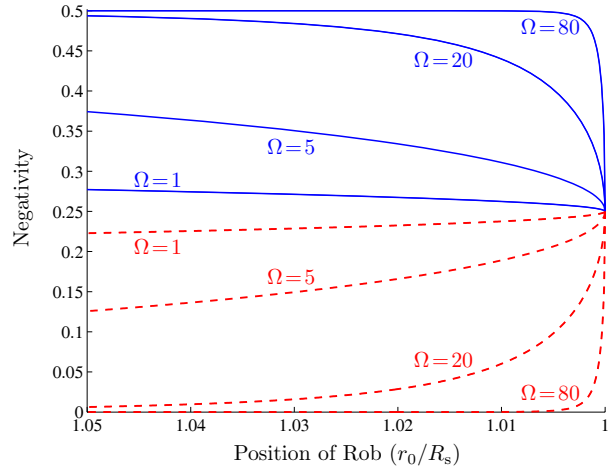}
\caption{Dirac field: Entanglement Alice-Rob (blue solid line) and
Alice-AntiRob (red dashed line). Universal conservation law for fermions
is shown for different values of $\Omega$. The entanglement
degradation in AR is quicker when $\Omega$ is smaller. The maximum
degradation is not maximal and its value is independent of $\Omega$.
}
\label{dncca}
\end{center}
\end{figure}

\begin{figure}
\begin{center}
\includegraphics[width=.50\textwidth]{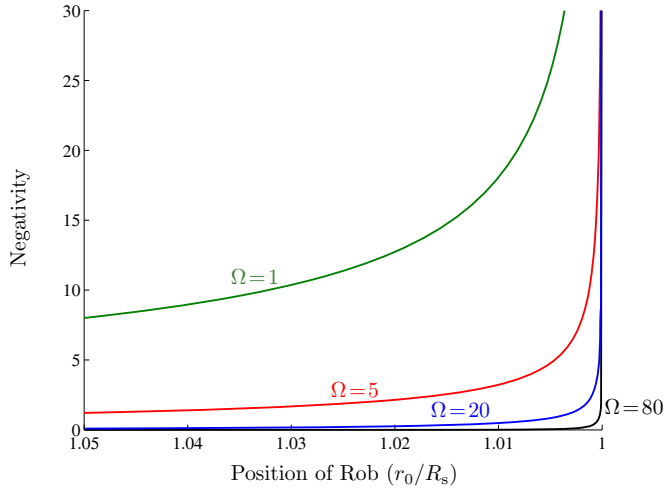}
\caption{Scalar field: Entanglement of the system Rob-AntiRob
(entanglement across the horizon) as a function of the position of
Rob for different values of $\Omega$. Entanglement diverges as Rob
approaches   the Schwarzschild radius.}
\label{snrar}
\end{center}
\end{figure}

\begin{figure}
\begin{center}
\includegraphics[width=.50\textwidth]{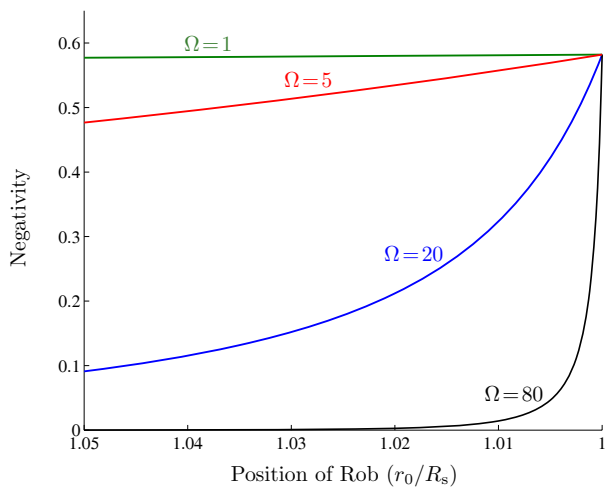}
\caption{Dirac field: Entanglement of the system Rob-AntiRob
(entanglement across the horizon) as a function of the position of Rob for
different values of $\Omega$. Entanglement tends to a finite value as Rob
approaches   the Schwarzschild radius.}
\label{dnrar}
\end{center}
\end{figure}

Hence, to compute it, we will need the partial transpose of the bipartite
density matrices \eqref{traza2}. The details associated to the
diagonalization of the partial transposed density matrices for each
subsystem are technically very similar to the Rindler case, and are not of
much interest for the purposes of this article. All the technical aspects of
such calculations can be found in \cite{Edu4} for Dirac and scalar fields.
The results of those calculations are shown in Figs. \ref{sncca} to
\ref{dnrar}. In Figs. \ref{sncca} and \ref{dncca} we can see the behavior
of the negativity on the CCA bipartitions for different values of Rob's
frequency $\Omega$.

For the scalar field we can see that as Rob is closer to the event horizon
the entanglement shared between Alice and Rob decreases. In the limit in
which Rob is very close to the horizon, entanglement is completely lost.
With the study performed in this work we can see the functional
dependence of the entanglement with the distance to the horizon. As seen in the figures,
 the degradation phenomenon occurs in a narrow region very close
to the event horizon. If Rob is far enough from the black hole he will not
appreciate any entanglement degradation effects unless either the mass of the black hole or the frequency of the mode considered are extremely small. There must be, indeed, a minimum residual effect associated to the Hawking thermal bath experienced in the asymptotically flat region of the spacetime, far from the region in which this approximation is valid, but it is unnoticeable small. Certainly, as it will be seen in fig. \ref{realb} and the discussion below, even very close to the horizon no effective entanglement degradation 
 occurs for physically meaningful values of mass and frequency.  
 
If we keep the frequency measured by Rob $\omegar$ constant,
$\Omega$ will grow proportional to  the black hole mass. With this in
mind, Fig. \ref{sncca} shows that the degradation is stronger for less
massive black holes. This result is consistent with the fact that the
Hawking temperature increases as the mass of the black hole goes to
zero. In the next section (specifically in Fig. \ref{realb}) we will show
that this is not an effect of choosing natural units, when an observer is at
a fixed distance of a black hole, the degradation will be higher for less
massive black holes.

In any case, for the scalar field, the entanglement in the system AR is
completely degraded when one of the observers is resisting very close to
the event horizon of the black hole. Hence, in this scenario, no quantum
information resources can be used (for instance to perform quantum
teleportation or quantum computing) between a free-falling observer and
an observer arbitrarily close to an event horizon. Moreover, no
entanglement of any kind is created among the CCA bipartitions of the
system (the ones who can classically communicate). Therefore, all useful
quantum correlations between a free-falling observer and an observer at
the event horizon are lost due to the Hawking effect degrading all the
entanglement in the system.

For the Dirac field  (Fig. \ref{dncca}) something very different happens.
We see that correlations in the bipartition AR decrease to a certain
finite limit, which means that there is entanglement survival even when
Rob is asymptotically close to the event horizon. This survival is a well
known phenomenon in the Rindler case \cite{AlsingSchul,Edu2}. At the
same time that entanglement is destroyed in the AR bipartition,
entanglement is created in the complementary $\text{A}\bar{\text{R}}$
bipartition so that negativity in the CCA bipartitions fulfills a
conservation law regardless of the distance to the event horizon and the
mass of the black hole
\begin{equation}\label{conservan}
\mathcal{N}_\text{AR}+\mathcal{N}_{\text{A}\bar{\text{R}}}=\frac12.
\end{equation}
The nature of this entanglement and the survival of correlations, even in
the limit of positions arbitrarily close to the horizon, is discussed in
\cite{Edu4,Edu5} for the Rindler case. When we deal with fermionic
fields there are correlations that come from the statistical fermionic
nature of the field which we cannot get rid of. The hypothesis is that this
entanglement, which is purely statistical, is the second quantized version
of the statistical entanglement disclosed in \cite{sta1}. Here we see that
the same conclusions drawn in that case can be perfectly applied to the
Schwarzschild black hole case.

About the dependence of the entanglement degradation on the frequency
of the Boulware mode, Fig. \ref{sncca}  shows that, for a scalar field, the
loss of entanglement between a free falling observer and an observer
outside but very close to the event horizon (AR) is greater for modes of
lower frequency. This makes sense because, energetically speaking, it is
cheaper to excite those modes and, therefore, they are more sensitive to
the Hawking thermal noise. For a Dirac field (Fig. \ref{dncca}) we see a
similar behavior, namely, lower frequencies are less protected against
entanglement degradation due to Hawking effect. However, the surviving
entanglement in the limit in which Rob is infinitely close to the event
horizon is not sensitive to the frequency of the mode considered;
remarkably,  the entanglement decays up to the same finite value for all
modes. This is in line with the idea that the entanglement that survives the
event horizon is merely due to statistical correlations,  and the only
information that survives when Rob is exactly at the horizon is the fact
that the field is fermionic as it is discussed in \cite{Edu3,Edu4,Edu5}.

From Figs. \ref{sncca} and \ref{dncca} we can also conclude that all the
relevant entanglement degradation phenomena is produced in the
proximities of the event horizon so that the Rindler approximation that
we are carrying out is valid  (Eq. \ref{conditioaproximatio2}). We can also
see that the degradation is small even in regions in which the
approximation  still holds. Therefore for longer distances from the
horizon the presence of event horizons is not expected to perturb
entangled systems.

In Figs. \ref{snrar} and \ref{dnrar} we can see the behavior of the
negativity on the $\text{R}\bar{\text{R}}$ bipartition for scalar and
Dirac fields respectively. Here we see that quantum correlations across
the horizon are created as Rob is standing closer to the event horizon. In
other words,  as Rob is getting closer to the event horizon the partial
system $\text{R}\bar{\text{R}}$ gains quantum correlations. This result
shows that, when Rob is near the horizon, the field states in both sides of
the event horizon are not completely independent. Instead, they get more
and more correlated. However, this $\text{R}\bar{\text{R}}$
entanglement is useless for quantum information tasks because classical
communication between both sides of an event horizon is forbidden, so no
quantum teleportation can be performed across the event horizon. It is
well known for the Rindler case that quantum correlations are created
between Rob and AntiRob \cite{Edu4} when the acceleration increases.
Here we see the direct translation to the Kruskal scenario. The growth of
those correlations encodes information about the dimension of the Fock
space for each field mode \cite{Edu5}.

%\begin{figure}[h]
%\begin{center}
%\includegraphics[width=.50\textwidth]{arbosw}
%\caption{(Scalar field: Entanglement of the system Alice-Rob as a function of the position of Rob for different $\omega$. Entanglement vanishes as Rob %approaches to the Schwarzschild radius. Black hole mass is fixed $m=1$}
%\label{Kruskal}
%\end{center}
%\end{figure}

\subsubsection{Mutual information}

\begin{figure}
\begin{center}
\includegraphics[width=.50\textwidth]{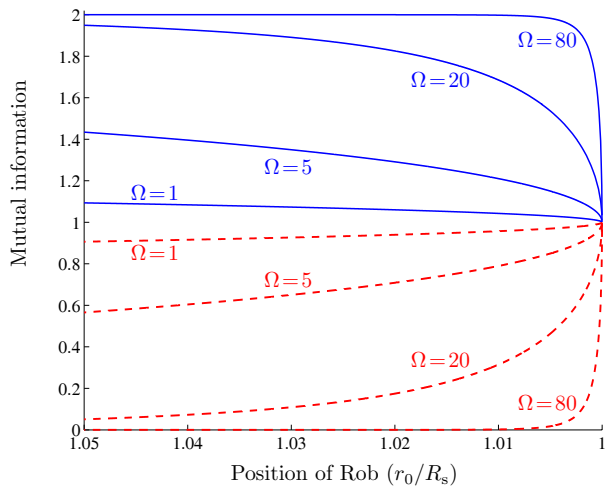}
\caption{Scalar field: Mutual information Alice-Rob (blue solid line) and
Alice-AntiRob (red dashed line). A conservation law (derived from the
behavior of purely classical correlations) is shown.  Mutual information  AR
decreases as Rob is closer to the horizon and mutual information
$\text{A}\bar{\text{R}}$ grows.}
\label{smi}
\end{center}
\end{figure}

\begin{figure}
\begin{center}
\includegraphics[width=.50\textwidth]{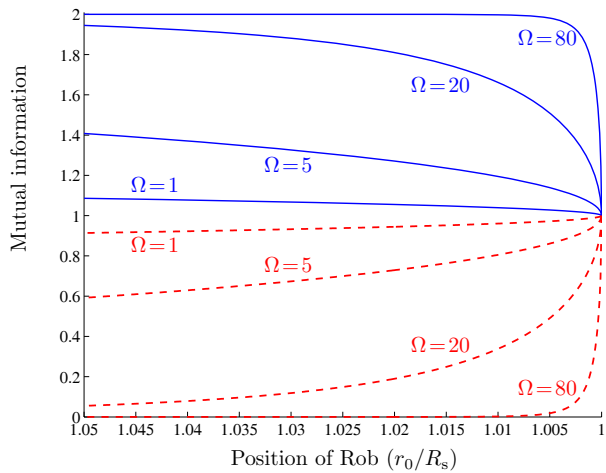}
\caption{Dirac field: Mutual information Alice-Rob (blue solid line) and
Alice-AntiRob (red dashed line). A conservation law (derived from the
behavior of purely quantum correlations) is shown.  Mutual information  AR
 decreases as Rob is closer to the horizon and mutual information $\text{A}\bar{\text{R}}$
  grows.}
\label{dmi}
\end{center}
\end{figure}

To account for all the correlations among the different bipartitions of
the system we will use the mutual information, which accounts for
correlations (both quantum and classical) between two different parts of
a system. For a bipartite system $\text{AB}$, it is defined as
\begin{equation}\label{mutualdef}
I_{\text{AB}}=S_\text{A}+S_\text{B}-S_{\text{AB}},
\end{equation}
where $S_\text{A}$, $S_\text{B}$ and $S_{\text{AB}}$ are respectively the Von Neumann
entropies $S =\tr\left(\rho \log_2\rho \right)$ for the individual
subsystems $\text{A}$ and $\text{B}$ and for the joint system $\text{AB}$. To compute the
mutual information  for each bipartition we will need the eigenvalues of
the corresponding density matrices. Again the technicalities of this
analysis can be found elsewhere \cite{Edu2,Edu4}. The results for the
CCA bipartitions are shown in Figs. \ref{smi} and \ref{dmi}.

We see here that we obtain the black hole version of the mutual
information universal conservation law found in previous works for the
Rindler case \cite{Edu4}. Namely, for any distance to the horizon or
black hole mass it is fulfilled that
\begin{equation}\label{conservami}
I_\text{AR}+I_{\text{A}\bar{\text{R}}}=2.
\end{equation}
Although, as we can see comparing Figs. \ref{smi} and  \ref{dmi}, the
behavior of the mutual information is very similar for both fermions and
bosons, the origin of this conservation law near the event horizon is
completely different.

For scalar fields this conservation near the horizon responds to a
conservation of  classical correlations only. This can be deduced from Fig.
\ref{sncca} which shows that quantum correlations drop very quickly as
the distance of Rob to the horizon decreases and, consequently, the only
correlations left  must be classical. However, the conservation of
classical correlations in the CCA bipartitions has to do with the
infiniteness of the dimension of the Hilbert space, as it is shown in
\cite{Edu5}. If the dimension of a bosonic field is limited to a finite value,
classical correlations also drop  as Rob is closer to the horizon (as
quantum correlations do).

On the other hand, a Dirac field has a built-in dimensional limit for the
Hilbert space of each mode imposed by Pauli exclusion principle. Although
previous works demonstrated that this limit in the dimension has nothing
to do with the behavior of quantum correlations \cite{Edu3,Edu4}, it
does limit the creation of classical correlations. Analogously to what is
discussed in \cite{Edu5}, the origin for the conservation law
(\ref{conservami}) in the fermionic case is a direct consequence of the
quantum correlations conservation law \eqref{conservan}.

The conclusion is that, although \eqref{conservami} is universal for scalar
and Dirac fields in the proximity of an eternal black hole, its origin is
completely different. For scalar fields it responds to a conservation of
classical correlations while for Dirac fields it is reflecting the quantum
correlations conservation \eqref{conservan}.

Mutual information for the $\text{R}\bar{\text{R}}$ bipartition does
not add any new result as it inherits the quantum correlations behavior
showed in Figs. \ref{sncca} and \ref{dncca}.

\subsection{Entanglement degradation dependence on the black hole mass}

In this section we will analyze the entanglement degradation for an
observer with the same characteristics in the presence of  different
black holes. To do so we are going to use the full dimensional quantities
$\omegar$ and $\Delta_0$.

We will consider that Rob's mode frequency is $\omegar=1.5$ Mhz, and
he is standing at a distance $\Delta_0=1$ cm and $\Delta_0=10$ cm
from the event horizon of  black holes with different masses, while he
shares an entangled state
\eqref{Haw1} or \eqref{Hawf} with a free-falling observer Alice.

\begin{figure}
\begin{center}
\includegraphics[width=.50\textwidth]{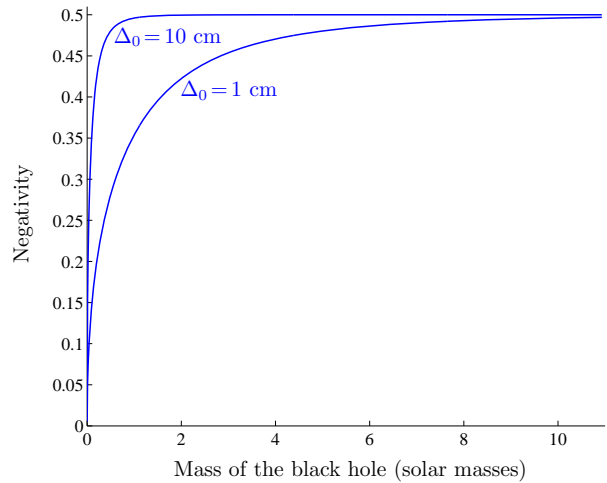}
\caption{Scalar field: Entanglement Alice-Rob when Rob stands at a distance
of 1 cm and 10 cm from the event horizon for a fixed frequency
$\omegar=1.5$ Mhz as a function of the black hole mass. Notice that,
for these values of $\Delta_0$, the approximation holds perfectly for any mass $m>10^{-5}$ solar masses.}
\label{realb}
\end{center}
\end{figure}

\begin{figure}
\begin{center}
\includegraphics[width=.50\textwidth]{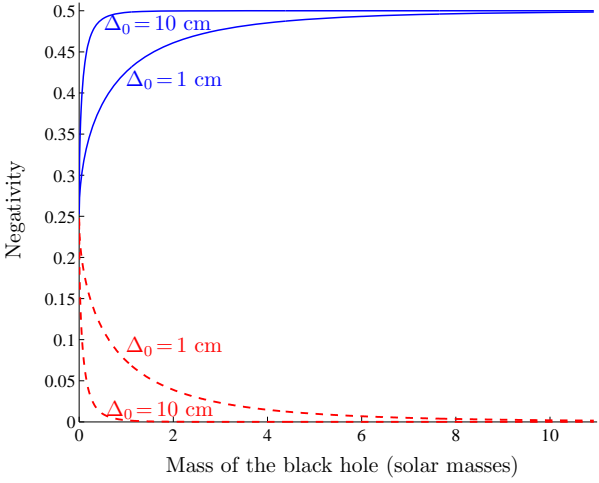}
\caption{Dirac field: Entanglement AR (blue continuous line) and
 $\text{A}\bar{\text{R}}$ (red dashed line) when Rob stands at a
 distance of 1 cm and 10 cm from the event horizon for a fixed frequency
  $\omegar=1.5$ Mhz as a function of the black hole mass.
  Notice that, for these values of $\Delta_0$, the approximation holds perfectly for any mass $m>10^{-5}$ solar masses.}
\label{realf}
\end{center}
\end{figure}

The quantum correlations that Rob and Alice share are shown in Figs.
\ref{realb} and \ref{realf} for scalar and Dirac fields, respectively.
From these figures we see that for a really close distance from the event
horizon, only small black holes would produce significant entanglement
degradation. Actually, the degradation decreases very quickly as the
black hole mass is increased.

%\begin{figure}
%\begin{center}
%\includegraphics[width=.50\textwidth]{fm4bos}
%\caption{(Scalar field: Entanglement of the system Alice-Rob as a function of the position of Rob for different frequencies $\Omega$. Entanglement vanishes as Rob approaches to the Schwarzschild radius. Universal behavior for all black holes}
%\label{suniv}
%\end{center}
%\end{figure}
%\begin{figure}
%\begin{center}
%\includegraphics[width=.50\textwidth]{f10mfer}
%\caption{(Diracr field: Entanglement of the system Alice-Rob as a function of the position of Rob for different frequencies $\Omega$. Entanglement vanishes as Rob approaches to the Schwarzschild radius. Universal behavior for all black holes}
%\label{duniv}
%\end{center}
%\end{figure}

Furthermore, we can see that the effects on the entanglement decrease
very quickly as the distance to the event horizon is increased. This shows
that quantum information tasks can be safely performed in universes that
present event horizons since only in the closest vicinity of the less massive
black holes the Hawking effect impedes the application of quantum
information protocols.

\section{Localization of the states}\label{newsec5}

Along this work we have used a plane-wave-like basis to express the quantum state of the field for the inertial an accelerated observers. These plane wave modes are completely delocalized, and therefore, they are not the most natural election of modes if we want to think of the observers Alice and Rob spatially localized to some degree.

However, a very similar analysis to the one carried out in sections \ref{sec2} and \ref{sec3} can be performed using a complete set of wave packet modes for both the Minkowski and Rindler solutions of the wave equation. These modes can be spatially localized and provide a clearer physical interpretation for Alice and Rob, which will eventually have to carry out measurements on the field. The way to build these wave packet modes can be found elsewhere \cite{Hawking0,Takagi,mods}.

The elements of this basis are defined as a function of the plane wave modes \eqref{modmin} as
\begin{equation}
u^{\text{M}}_{\hat\omega,l}=\frac{1}{\sqrt{\epsilon}}\int_{\hat \omega}^{\hat \omega+\epsilon}d\nu\, e^{-i \nu l}u^{\text{M}}_{\nu},
\end{equation}
where $\hat\omega$ and $l$ label each wave packet.
 
 We can define creation an annihilation operators associated to these wavepackets $a_{\hat \omega,l,\text{M}}$, $a_{\hat\omega,l,\text{M}}^\dagger$  such that $a_{\hat\omega,l,\text{M}}$ annihilates the Minkowski vacuum and $a_{\hat\omega,l,\text{M}}^\dagger \ket{0}_\text{M}=\ket{1_{\hat\omega,l}}_\text{M}$ represents a wavepacket peaked for a frequency $\hat \omega$ and whose spatial localization can be associated to the maximum of $u^{\text{M}}_{\omega,l}$ as a function of $\hat x$ and $\hat t$.
 
A similar analysis can be done for the Rindler basis
\begin{align}
\nonumber u^{\text{I}}_{\omega,l'}&=\frac{1}{\sqrt{\epsilon}}\int_{\omega}^{ \omega+\epsilon}d\nu\, e^{-i \nu l'}u^{\text{I}}_{\nu},\\
u^{\text{IV}}_{\omega,l'}&=\frac{1}{\sqrt{\epsilon}}\int_{\omega}^{ \omega+\epsilon}d\nu\, e^{-i \nu l'}u^{\text{IV}}_{\nu},
\end{align}
 $\omega$ and $l'$ label each wave packet.  We can define creation an annihilation operators associated to these wavepackets $a_{\omega,l',R}$, $a_{\omega,l',R}^\dagger$ (where $R=\text{I},\text{IV}$) such that $a_{\omega,l',R}$ annihilates the region Rindler region $R$ vacuum and $a_{\omega,l',R}^\dagger \ket{0}_R=\ket{1_{\omega,l'}}_R$ represents a wavepacket peaked for a Rindler frequency $\omega$ and whose spatial localization can be associated to the maximum of $u^R_{\omega,l'}$ as a function of $x$ and $t$.
 
 We can compute then the Bogoliuvob transformation between the Minkowski wavepackets and the Rindler wavepackets \cite{mods}
 \begin{equation}
a_{\hat\omega,l,\text{M}}=\alpha^{\text{I}*}_{\omega,l',\hat \omega,l}\,
a^{\phantom{\dagger}}_{\omega,l',\text{I}}+\beta^{\text{IV}}_{\omega,l',\hat \omega,l}\,
a^\dagger_{\omega,l',\text{IV}}.
\end{equation}
Where the Bogoliuvob coefficients are computed in the same fashion as for the plane wave case
\begin{equation}
\alpha^{\text{I}}_{\omega,l',\hat \omega,l}=\left(u^{\text{I}}_{\omega,l'},u^{\text{M}}_{\hat\omega,l}\right),
\quad\beta^{\text{IV}}_{\omega,l',\hat \omega,l}=-\left(u^{\text{IV}}_{\omega,l'},u^{\text{M}*}_{\hat\omega,l}\right).
\end{equation}

It is shown in \cite{mods} that, apart from an irrelevant phase factor, the Bogoliuvob coefficients are related with \eqref{bogo1} as follows
\begin{align}\label{bogonew}
\nonumber\alpha^{\text{I}}_{\omega,l',\hat \omega,l}&=\hat\alpha^{\text{I}}_{ij}\,\mathcal{G}_\alpha(\hat\omega,l,\omega,l'),\\
\beta^{\text{I}}_{\omega,l',\hat \omega,l}&=\hat\beta^{\text{IV}}_{ij}\,\mathcal{G}_\beta(\hat\omega,l,\omega,l').
\end{align}
It is shown in \cite{mods} that $\mathcal{G}_\alpha(\hat\omega,l,\omega,l')\approx \delta_{\omega\omega_\alpha}\delta_{ll_\alpha}$ and  $\mathcal{G}_\beta(\hat\omega,l,\omega,l')\approx \delta_{\omega\omega_\beta}\delta_{ll_\beta}$, where $l_\alpha=l_\alpha(l')$ and $\omega_\alpha =\omega_\alpha( \omega,l')$.

The key feature of this transformations is that they have again a diagonal form. As it can be read from \eqref{bogonew}, a Minkowski wavepacket $\ket{1_{\omega,l'}}_\text{M}$ is connected with a pair of Rindler wavepackets in regions I and IV. Moreover, the functional form of the dependence of this coefficients with the acceleration is effectively the same. This analysis made for the Rindler and Minkowskian modes can be straightforwardly translated to the Boulware and Hartle-Hawking modes.  A completely analogous analysis can be done for the fermionic case.

Consequently all the conclusions extracted in this article for delocalised modes are also valid for the localized modes defined above.

\section{Conclusions}\label{conclusions}

We have analyzed the entanglement degradation produced in the vicinity
of a Schwarzschild black hole.

With this aim, we have carried out a detailed study of the Schwarzschild
metric in the proximity of the horizon, showing how we can adapt the
tools developed in the study of the entanglement degradation for
uniformly accelerated observers \cite{Alicefalls,AlsingSchul,Edu2,Edu4}
to the black hole case. In particular, we have shown that, regarding
entanglement degradation effects, the Rindler limit of infinite acceleration reproduces
a black hole scenario in which Rob is arbitrarily close to the event horizon.
More importantly, we have shown the fine structure of this  limit, making
explicit the dependence of the entanglement degradation phenomena on the distance
to the horizon, the mass of the black hole, and the Boulware frequency
$\omegar$ of the entangled mode under consideration, while keeping
control of the approximation to make sure that the toolbox developed for
the Rindler case can be still rigorously used here.

By means of this analysis we have seen that all the interesting
 entanglement degradation phenomena due to the Hawking effect are produced very
close to the event horizon of the Schwarzschild black hole. The
entanglement degradation introduced by the Hawking effect becomes
quickly negligible as Rob is further away from the event horizon. In  other
words, quantum information tasks done far away from event horizons are
not perturbed by the existence of such horizons.

We have also shown that for a fixed Rob's mode frequency and at a fixed
distance from the event horizon the entanglement degradation is greater
for less massive black holes. This is consistent with the fact that the
Hawking temperature is higher for less massive black holes. Furthermore,
  the Hawking entanglement degradation is a universal
phenomenon in the sense that the degradation depends only on Rob's
frequency and his distance to   the horizon in units natural to the black
hole (namely, the surface gravity for frequencies and the Schwarzschild
radius for distances). In these units, there is no extra dependence on the
black hole mass,  as expected.

We have been able to adapt all the conclusions drawn for the Rindler case
to the Schwarzschild scenario. In particular, we have seen that bosonic
and fermionic entanglement behave in a very different way in the
proximity of a black hole. As it was known for the Rindler case
\cite{Edu4}, entanglement on the CCA bipartitions is completely lost for
the scalar field while there is a quantum correlation  conservation law for
the Dirac field.

In \cite{Edu3} it was shown that for two different kinds of fermionic
fields (Dirac fields or Grassmann scalars) and also for different
maximally entangled states (occupation number or spin Bell states) the
entanglement in the CCA bipartitions behaves exactly the same way. This
fact was used to argue that  it is statistics and not dimensionality that
determines  the behavior  of correlations in the CCA bipartitions in the
case of uniformly accelerated observers. This study
proves that this argument is also valid for Schwarzschild black holes, not
only in the limit in which Rob is on the event horizon, but in the whole
region in which the interesting entanglement degradation phenomena are produced.
Therefore, the universal fermionic entanglement behavior is also manifest
in the presence of a black hole.

For the Schwarzschild case, there also appears the  universal mutual
information conservation law found for both scalar and Dirac fields in
the Rindler case \cite{Edu4}. In the fermionic case, it is due to a
conservation of quantum correlations while, for bosons, it only reflects
the conservation of classical correlations  that happens in the case of
infinite dimensional Hilbert spaces for each mode.

Moreover,   as Rob is getting closer to the event horizon, quantum
correlations between modes on both sides of the event horizon are
created, namely the correlations between field modes in region I and IV
of the Kruskal space-time grow up to a value determined by the dimension
of the Hilbert space of each mode, which is finite for the fermionic case
and infinite for the scalar field.

 As discussed previously for the Rindler scenario \cite{Edu3,Edu4,Edu5}, and unveiled here for the Schwarzschild black hole case, under the hypothesis that the fermionic entanglement which survives the event horizon is of statistic nature as in \cite{sta1}, it would not contain any useful information. Therefore if an entangled pair is created close to the event horizon (for instance particle/antiparticle creation) and one of the subsystems falls into the black hole while the other resists close to the horizon, no other entanglement except for the mere statistical would survive the degradation provoked by the Hawking effect.

The problem of the localization of the Rindler and Minkowski modes has also been analyzed, showing that the results obtained here can be extrapolated to the case in which we consider a complete set of localized wave packets as a basis of the Fock space for the inertial and accelerated observers.

The scenario that we have discussed here is that of a static eternal black
hole in which no dynamics is present. The analysis of entanglement
degradation due to the dynamical creation of an event horizon in a
gravitational collapse scenario is under current development and will be
reported elsewhere.

\section{Acknowledgments}

The authors would like to thank Bei-Lok Hu, Ivette Fuentes, Robert Mann and Paul Alsing for the helpful discussions during the International Workshop on Relativisitc Quantum Information (RQI-N 2010).

The authors also thank Jorma Louko and Carlos Barcel\'o for their helpful comments and observations.

This work was partially supported by the Spanish MICINN Projects
FIS2008-05705/FIS and FIS2008-06078-C03-03, the CAM research
consortium QUITEMAD S2009/ESP-1594, and the Consolider-Ingenio
2010 Program CPAN (CSD2007-00042). E. M-M was partially supported
by a CSIC JAE-PREDOC2007 Grant.

\bibliographystyle{apsrev}

\end{document}